\documentclass[AMA,LATO1COL]{WileyNJD-v2} 

\articletype{Research Article}%

\usepackage{amssymb,amsfonts}
\usepackage{moreverb}
\usepackage{subfig}
\usepackage{siunitx}

\begin{document}
\title{Quasi-static closed-loop wind-farm control for combined power and fatigue optimization}

\author[1]{Ishaan Sood*}

\author[1]{Christophe del Fosse et d'Espierres}

\author[1]{Johan Meyers}

\authormark{SOOD \textsc{et al}}

\address{\orgdiv{Department of Mechanical Engineering}, \orgname{KU Leuven}, \orgaddress{\state{Leuven}, \country{Belgium}}}

\corres{\email{ishaan.sood@kuleuven.be}}

\abstract[Abstract]{To counteract detrimental turbine--turbine aerodynamic interactions within large farms and increase overall power production, closed-loop wind-farm control strategies such as wake steering have emerged as a popular means to facilitate real-time wind farm flow control. The optimal wake steering set points to maximize farm power production for a given inflow condition are generally determined using fast engineering models. However due to a lack of fast structural models, influence of wind farm flow control on turbine structural fatigue and loading is generally not considered. In this work, we develop a methodology for combined power and loads optimization by coupling a surrogate loads model with an analytical quasi-static Gaussian wake merging model. The look-up table based fatigue model is developed offline through a series of OpenFast simulations, covering different operational states of a DTU \SI{10}{\mega\watt} reference wind turbine, and verified against large eddy simulations with aeroelastic coupling. Subsequently, optimal control set points for the TotalControl reference wind power plant are obtained using the analytical model, and tested in a wind farm emulator that is based on large eddy simulations. The wake model is calibrated online using in a quasi-static closed-loop manner. Benefits of the closed-loop controller are exhibited via comparison of farm performance against greedy operation and against open-loop control results obtained without feedback or calibration. Results show that the closed-loop control out performs open-loop control, with farm configurations with deep turbine arrays showcasing the highest gains. Inclusion of fatigue in the cost function through the developed LUT also leads to interesting insights, with reduced blade root fatigue loading without significant decrease in power production when compared to open-loop control. A case study is also performed which showcases real world applications for the developed closed-loop controller and the load LUT, in which the closed-loop controller is shown to react to scenarios with turbine operation shut-down, optimizing the yaw angles online to maximize performance for the new layout.}


\maketitle

\section{Introduction}\label{sec1}

Modern mega-watt sized wind turbines are often placed in clusters called wind farms to reduce installation and operational costs. However due to their close arrangement, the wakes of upstream turbines tend to superimpose on downstream turbines leading to performance losses and fatigue degradation across the farm. To counteract this effect, recent literature has proposed wind-farm flow control (WFFC) strategies such as wake steering and induction control which can be employed together or individually, in either an open-loop or closed-loop control framework.~\cite{Munters2018, Debusscher2022, Meyers-2022} The current standard for WFFC is mostly open-loop control, validated through various recent field experiments.~\cite{Duc,Bossanyi_Ruisi, simley} The open-loop algorithm consists of fast wake models which are calibrated offline based on site specific conditions, and then used to develop optimal axial induction or wake steering set point look-up tables for a range of different atmospheric conditions such as wind speed, wind direction, turbulence intensity and stability. While efficient and simple to implement, the open-loop framework by definition has several drawbacks. The methodology does not involve any online model error corrections when compared to field measurements as the logic does not take into account real-time performance feedback from the wind farm, which can result in model prediction errors. Additionally, open-loop control strategies may be less effective for complex wind farms with significant variations in atmospheric conditions, and may be more susceptible to performance degradation over time as the wind farm and turbines age. Closed-loop control strategies, on the other hand, use real-time feedback from sensors to continuously adjust turbine settings to optimize performance. By continuously obtaining measurements from the wind farm, the underlying model used by the controller to predict farm performance can be calibrated online to adapt to the changing conditions in the field.~\cite{Doekemeijer2020, Sood_tuning}  Such frameworks also include real time estimators based on techniques such as Kalman filters or model inversions to improve the model's awareness of the current conditions in the farm and enable it to predict future conditions.~\cite{Doekemeijer_2018,9483088} The estimators rely on physical sensors such as lidars, or on virtual sensors which are capable of inferring the turbines operating states through easily measurable parameters such as rotor speed, power production, etc.

Additionally, based on the time-scales of the control action the control logic can be divided into two categories, quasi-static control and dynamic control. In quasi-static control, the control set points of the turbines are updated at time-scales comparable to the farm flow through time, reacting to slowly changing operational conditions such as change in wind speed or wind direction. Previous studies and commercial products have already utilized quasi-static control through static wake models to optimize wind farm performance and performed utility scale validation of the gains.~\cite{siemens_webpage,SoodAIAA,Fleming2019} Dynamic control on the other hand deals with changing control actions on the order of seconds, capable of reacting to fast changing conditions such as wind gusts and grid frequency demand. Dynamic control through large eddy simulations (LES) or dynamic wake models has shown significant wind farm performance gains in numerical environments. However this branch of WFFC is still in its infancy, with limited experimental validation.~\cite{Munters2018, frederik-2020} Further details of the different wind farm flow control strategies and their associated challenges have been compiled in recent reviews on the subject.~\cite{Meyers-2022, Houck2022, shapiro2022}

Wind farm flow control can be utilized to achieve a number of objectives, such as reducing Levelized Cost of Energy (LCOE) and increasing revenue,~\cite{Riva2019} mitigating structural loads~\cite{Vali2022} or provision of ancillary services. ~\cite{Shapiro2019} While WFFC has been predominantly focused on boosting farm wide power production, recent studies have shown that focusing solely on power optimization could lead to increase in turbine fatigue across the wind farm.~\cite{SoodAIAA} The type of control strategy chosen can also have significant impact on the turbine structural integrity of upstream and downstream turbines due to changing the turbine induction through pitch and tip speed ratio (TSR) control, or yawing the turbines to steer its wake. For instance while yaw control is the most investigated and developed coordinated control strategy for power maximization with commercially available solutions,~\cite{siemens_webpage} its influence on structural loading may be beneficial or detrimental based on the resulting partial wake for downstream turbines or the inflow operating conditions.~\cite{Herges2018, Damiani2018} While numerous studies have been performed on wake steering for power maximization, limited literature is available on combining turbine component loading in the optimization procedure. Thus, a significant open challenge in WFFC is the development of strategies which perform combined power and loads optimization. While accurate turbine structural analysis models exist,~\cite{Jonkman2018,OpenFAST2021} their computational costs are significantly greater than that of analytical wake models, making their use prohibitive for online optimization of the performance of large wind farms. Therefore, fast and accurate turbine loading models are needed which can be used along with wake models in the wind farm performance optimization framework.

\begin{figure}
    \centering
    \includegraphics[width=5in]{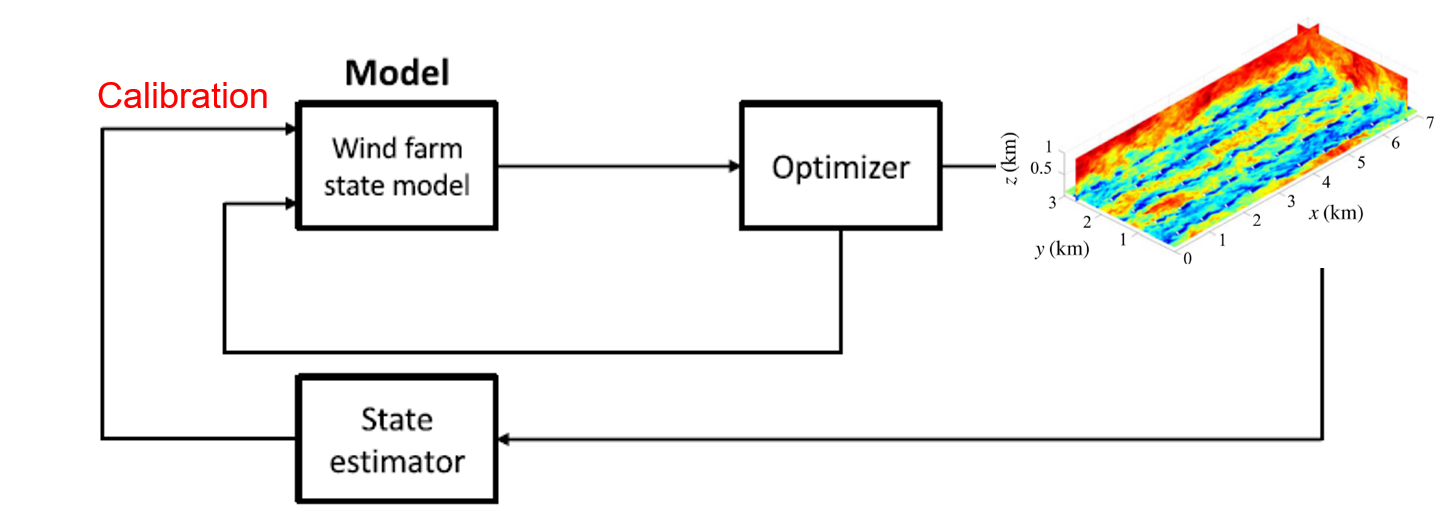}
    \caption{closed-loop wind farm control methodology. Calibration is required
based on measurements from a plant to improve the analytical wake model. A LES based virtual wind farm emulator is used as a substitute for real operation.}
    \label{fig:closed_loop}
\end{figure}

Hence in this work, we develop a quasi-static closed-loop wind-farm control framework to perform simultaneous power and fatigue optimization of a virtual wind farm. An overview of the closed-loop methodology is shown in Figure \ref{fig:closed_loop}. The analytical wind farm model used for determining optimal set points comprises of a Gaussian wake model using a recursive wake merging method, combined with a fatigue look-up table (LUT) developed offline for the DTU \SI{10}{\mega\watt} turbine using OpenFast.\cite{delFosseetdEspierresChristophe2021Doal,Lanzilao2020}. As a substitute for a real wind farm, a LES suite provides the virtual test bed to test the optimal set points, and provide feedback for the calibration of the state model. A database comprising of greedy operation and open-loop control, developed using the same LES suite for a reference wind farm, serves as a benchmark to quantify the benefits of the developed closed-loop controller. The reference simulations are also used for verifying the developed load LUT, before it is used for combined power and load optimizations.

The current work is organised as follows: first, the flow of information and different aspects of the closed-loop control framework are detailed, which include the wake model, the optimization methodology, the LES environment, the reference database and the online model calibration and estimation methodologies. The next section details the development and validation of a fatigue look-up table for the DTU \SI{10}{\mega\watt} turbine. This is followed by details of the optimization cases studied in this work. Results of the simulations and a comparison between greedy, open-loop and closed-loop control are presented in the next section on the basis of power production, controller dynamics and fatigue accumulation in the turbines. Case studies showcasing the real world application of the developed tools are presented in the next section. Finally, the work concludes with a discussion on the obtained results and suggestions for future work.

\section{Quasi-static closed-loop control methodology}

\begin{figure}
    \centering
    \includegraphics[trim=0 50 0 50,clip,width=6in]{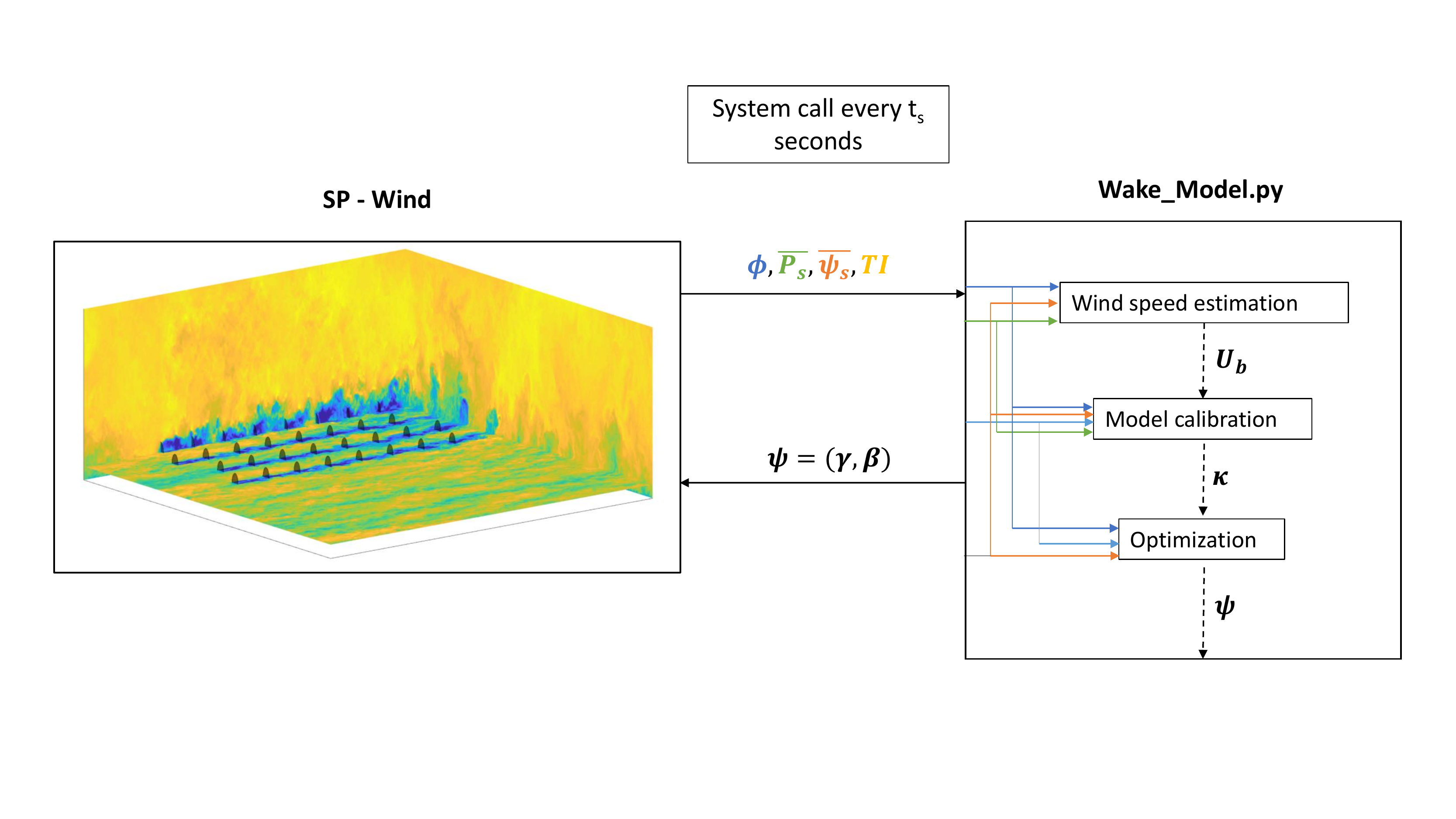}
    \caption{Flow of information in the quasi-static closed-loop control framework.}
    \label{fig:closed_loop_formulation}
\end{figure}

The formulation of the closed-loop control framework is summarized in Figure \ref{fig:closed_loop_formulation}, and the following sub-sections elaborate upon the different modules of the framework. In the current work, SP-Wind, a LES environment comprising of an aeroelastic wind turbine representation, is used as a substitute virtual testing environment for a real wind farm. In that sense, only the measurements that would be conventionally available from a wind farm through SCADA channels are considered, without assuming the installation of additional sensors such as lidars. Furthermore, since the wake model used in this work makes a steady state assumption, dynamically changing inflow conditions are accounted for in a quasi-static sense by executing the closed-loop framework at regular fixed intervals of $t_s$. At each optimization step, the Fortran based LES code SP-Wind calls the python based analytical wake model through system calls. Time averaged measurements over the last $t_s$ seconds are then taken from the wind farm operating in the LES domain, which are used to first estimate the background wind speed $U_b$, followed by the calibration of model parameters $\mathbf{\kappa}$, and then finally the optimization of wind farm performance. The turbine control parameters are designated by $\Psi$, and include the optimal set point for all the turbines within the farm to achieve the optimization objective. In this work, we restrict ourselves to wake steering control, but the framework could be extending to induction control through pitching the turbine blades as well. The choice of sampling time $t_s$ can have a major impact on the performance of the framework and the resulting wind farm performance, but as a first step it is set to \SI{600}{s}. This is chosen as a compromise between having enough averaging time to allow upstream control effects to advect downstream, and having sufficient update steps within the simulation time of \SI{3600}{s}, which is restricted per case due to computational resources. In practice, the effect of atmospheric turbulence, turbine spacing and yaw duty cycle fatigue should also be included when determining the sampling time,~\cite{Howland_yaw_22} but is not considered in the current work due to simulation time restrictions. Once the optimal set points are determined, they are sent back to SP-Wind. The turbines in the LES domain track the dynamically changing set points based on the yaw rate limitation of the turbines, which is set to \SI{10}{\deg\per\second} for the DTU\SI{10}{\mega\watt} turbine that we use in the current work. The optimal yaw angles are tracked using a simple on-off control logic.

\subsection{Wake Model}{\label{wake_model}}
 
In the developed quasi-static closed-loop control framework, an analytical wake model serves as the state model for the wind farm operation which is used for optimizing performance. The wake function is defined as per a new coordinate system, $\boldsymbol{X_i}(\boldsymbol{x})=(X_i(\boldsymbol{x}),Y_i(\boldsymbol{x}),Z_i(\boldsymbol{x}))$, for each turbine $i$. The wake deficit $W_i$ is then evaluated using the Bastankhah model,~\cite{Bastankhah2016} according to which the wake deficit behind a yawed turbine is evaluated as a function of stream-wise coordinates measured respect to turbine $i$ using the equation 
\begin{equation}
    W(\boldsymbol{x})) = \left(1-\sqrt{1-\frac{C_T cos\gamma}{\frac{8\sigma_x \sigma_y}{D^2}}})\right)exp\left[-\frac{1}{2}\left\{\left(\frac{z-z_h}{\sigma_z}\right)^2+\left(\frac{y-\delta}{\sigma_y}\right)^2\right\}\right],
\end{equation}
where $\gamma$ is the turbines yaw angle, \textit{$C_T$} is the wind turbine thrust coefficient, $\delta$ is the wake deflection and \textit{D} is the turbine diameter. Furthermore, $\sigma_y$ and $\sigma_z$ are wake widths of the turbine at the downstream location, which in turn depend upon the wake growth rate $k_w$ and the near-wake length $x_0$ according to the following expressions

\begin{equation}
    \frac{\sigma_y}{D} = 0.35 \cos{\gamma} + k_w \ln{\left[1+\exp\left(\frac{x-x_0}{D}\right)\right]},
\end{equation}

\begin{equation}
    \frac{\sigma_z}{D} = 0.35 + k_w \ln{\left[1+\exp\left(\frac{x-x_0}{D}\right)\right]}.
\end{equation}
Further details of the wake deficit model and its parameters can be found in the references~\cite{Bastankhah2016, zong_2020}. The different wakes within a farm are combined together using the Lanzilao recursive model~\cite{Lanzilao2020}. Assuming that the wakes in the wind farm are carried by the background flow $\boldsymbol{U}_b(\boldsymbol{x})$, the flow field in the farm is then given by the velocity field  $\boldsymbol{U}_{N_t}$, which is constructed using the recursive formula 
\begin{equation}
    \boldsymbol{U}_i(\boldsymbol{x}) = (\boldsymbol{U}_{i-1}(\boldsymbol{x})\cdot \boldsymbol{e}_{\bot,i})(1-W_i(\boldsymbol{x}))\boldsymbol{e}_{\bot,i} + (\boldsymbol{U}_{i-1}(\boldsymbol{x})\cdot \boldsymbol{e}_{\parallel,i})\boldsymbol{e}_{\parallel,i},\quad \text{for}\ i = 1,....,N_t
\end{equation}
The starting term of the recursion is given by $\boldsymbol{U}_o(\boldsymbol{x})=\boldsymbol{U}_b(\boldsymbol{x})$, which is an input to the model. Unit vectors $\boldsymbol{e}_{\bot,i}=(\cos\theta_i,\sin\theta_i)$ and $\boldsymbol{e}_{\parallel,i}=~(-\sin\theta_i,\cos\theta_i)$ account for the incoming wind direction and yaw angle at turbine $i$. The angle $\theta_i$ defines the orientation of turbine $i$ as 
\begin{equation}
    \theta_i = \arctan{V(\boldsymbol{x_i})/U(\boldsymbol{x_i}) + \gamma_i}.
\end{equation}
The inflow velocity of each turbine $i$ is finally evaluated across the disc area at observation points which are distributed across the disc according to a quadrature rule with $N_q$ = 16 points. The quadrature-point coordinates are denoted by $x_{k,q}$ and are chosen following the rule proposed by Holoborodko with uniform weighting factor of $w_q=1/N_q$ ~\cite{Pavel2011}. The inflow velocity at each turbine, $S_i$, is therefore calculated as

\begin{equation}
    S_i = \sum_{q=1}^{N_q}w_q S(x_{i,q}),
\end{equation}
where, $S(x)$  = $||U(x)||_2$. The inflow velocity at each turbine is then finally used to evaluate the power production of the turbine, through the expression $P = \frac{1}{2} \rho A C_P S^3$. In the power expression, $A$ is the area of the turbine and $C_P$ is the coefficient of power of each turbine, evaluated for a yaw angle $\gamma_k$ according to the cosine power law $C_P(\gamma)= C_P\cdot\cos(\gamma)^3$.\cite{burton2001wind} For the DTU \SI{10}{\mega\watt} turbine, previous research determined a scaling factor $\eta(\gamma) = 1.08/\cos{\gamma}$, to improve the power predictions while wake steering.\cite{Doekemeijer2020}

\subsection{Optimizer}

Turbine set point optimization is carried out by minimizing the following cost function,
\begin{equation}
    \min_{\mathbf{\gamma}} \quad - w_P P_{\textit{GAIN}} + w_L DEL_{\textit{GAIN}},
    \label{eq:optimization}
\end{equation}
where the terms $P_{\textit{GAIN}}$ and $DEL_{\textit{GAIN}}$ have the following definitions
\begin{equation}
    P_{\textit{GAIN}} = \sum_{k=1}^{N_t} \frac{C_P (\gamma_k) S_{k}^3 (\mathbf{\gamma})}{C_P (0) S_{k}^3 ({0})},
    \label{eq:pgain}
\end{equation}
\begin{equation}
    DEL_{\textit{GAIN}} = \sum_{k=1}^{N_t} \frac{DEL_k(S_k,\gamma_k)}{DEL_k(S_k,0)}.
\end{equation}
The first term in the cost function is the normalized power gain due to yaw control determined by the wake model when compared to a case without wake steering, with the constants such as density and turbine area cancelled out. The second term deals with the normalized fatigue load increase, in the form of blade root Damage Equivalent Loads (DEL) gains~\cite{Freebury2000}. The power production of each turbine depends upon the turbine inflow velocity $S_k$, while the resulting fatigue load per turbine is obtained via a fatigue LUT, which is defined in Section~\ref{fatigue_lut_section}. Weights $w_P$ and $w_L$ are added in the cost function to control the relative importance of the two terms, with $w_P + w_L = 1$ . The variable $\boldsymbol{\gamma}$ is a vector containing the yaw set points for all the turbines across the farm. Turbine yaw angles are limited between $\pm 30^{\circ}$ to prevent excessive fatigue. The optimisation problem is solved to obtain the yaw set points across the wind farm by using the SLSQP solver from the SciPy Python package, while utilizing the multi-start approach of basin-hopping to avoid local minima.\cite{2020SciPy-NMeth}

\subsection{LES environment and Reference cases}{\label{ref_database}}

\begin{figure}
    \centering
    \includegraphics[width=4in]{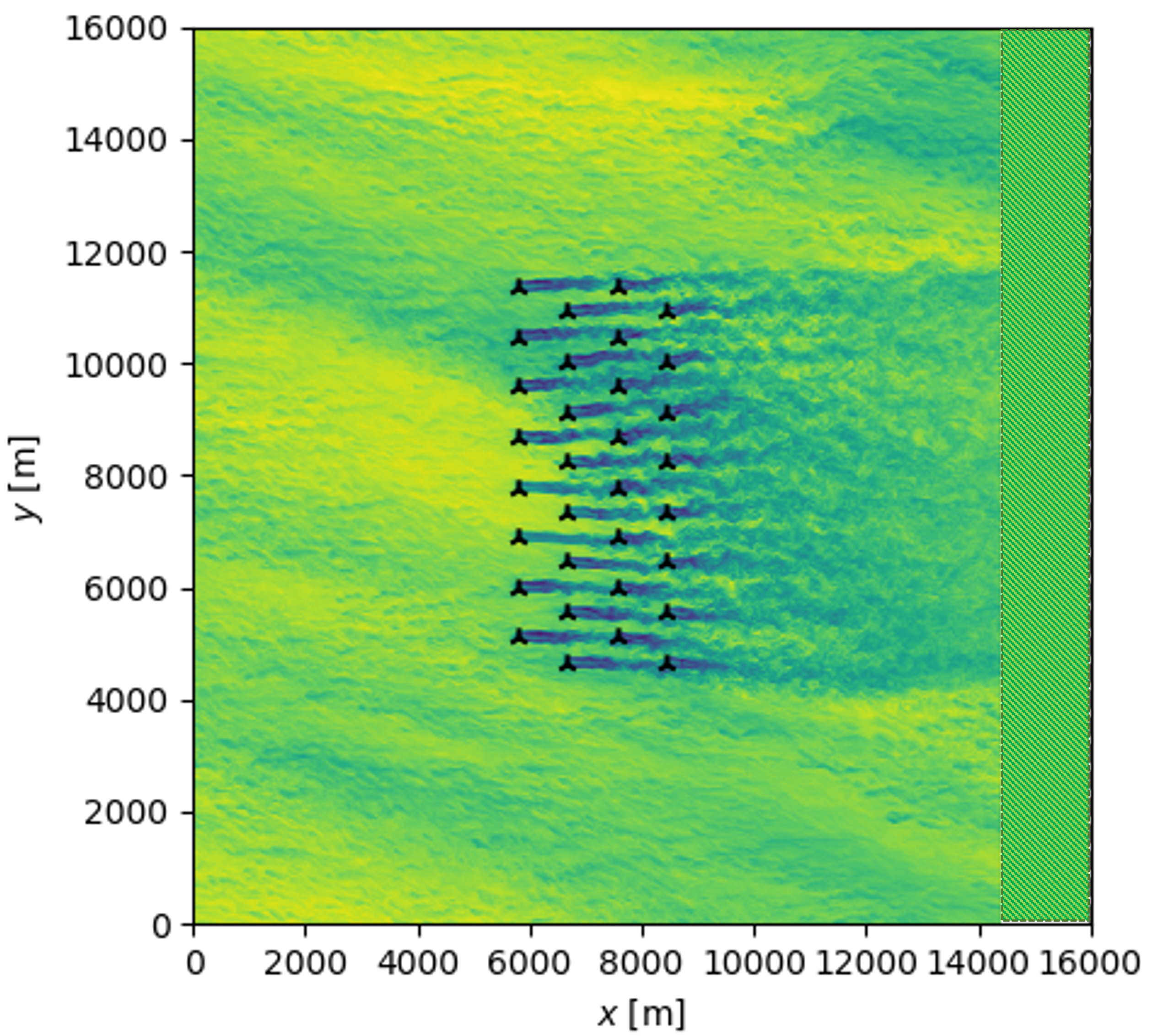}
    \caption{Layout of the Total Control reference wind farm comprising of 32 DTU \SI{10}{\mega\watt} turbines in the SP-Wind simulation domain. The hashed region in the last 10\% of the domain represents the fringe region. Turbines are numbered 1 to 32 starting at the bottom left of the farm, increasing left to right and bottom to top.}
    \label{fig:tc_rwp}
\end{figure}

Reference wind farm cases operating under normal operation, as well as under open-loop control are required to validate the developed look up table and also to demonstrate the benefits of closed-loop control. To this end, we make use of the publicly available TotalControl reference wind farm database which comprises of numerical measurements obtained from LES spanning two types of boundary layers, Pressure Driven Boundary Layers (PDBL) and Conventionally Neutral Boundary Layers (CNBL). In total, 5 different inflow boundary layers are used, 2 PDBL with different surface roughnesses, and 3 CNBL with different capping inversion strengths and boundary layer height. The wind farm used for generating the reference database is the TotalControl reference wind power plant (TC RWP), which comprises of 32 DTU \SI{10}{\mega\watt} turbines. Inflow conditions are then expanded by considering different wind directions through counter clockwise rotation of the entire wind farm in the simulation domain, thus resulting in an open-source database comprising 10 wind-farm simulations operating with greedy operation.~\cite{zenodo_pdk_0,zenodo_pdk_30,zenodo_pdk_90,zenodo_pdkhi_0,zenodo_cnk2_30,zenodo_cnk2_60,zenodo_cnk4_30,zenodo_cnk4_90,zenodo_cnk8_0,zenodo_cnk8_90} The reference database was created using SP-Wind, an in-house research code developed over the past 15 years for wind farm simulations, optimizations and field data validation studies.~\cite{Goit2015,Allaerts2018,Munters2016,sood_lillgrund} SP-Wind solves the filtered Navier--Stokes equations, with turbine forces being parameterized using an aeroelastic actuator sector model to allow for accurate turbine representation and time series analysis of turbine power and structural loading.~\cite{Vitsas2016} The concurrent precursor method is used to generate the turbulent inflow for the farm.~\cite{Stevens2014}  In this work, we use the same virtual environment and numerical set-up as the reference wind farm cases. The domain and discretization details are summarized in Table \ref{tab:spec_numerics}, and the layout of the wind farm in the domain is shown in Figure \ref{fig:tc_rwp}.
\begin{table}
    \caption{Summary of the general domain parameters}
    \centering
    \begin{tabular}{lcc}
    \hline
    Domain size & $L_x \times L_y \times L_z$ & $16 \times 16 \times 1.5$ \SI{}{km^3} \\
    Grid &$N_x \times N_y \times N_z$&$1200 \times 1200 \times 225\ $\\
    Resolution&$\Delta_x \times \Delta_y \times \Delta_z$&$13.33 \times 13.33 \times 6.66 \SI{}{m^3}$\\
    Wind-farm spin-up time&$T_{\textit{spin}}$&$900$ \SI{}{s}\\
    Simulation time&$T$&$3600$ \SI{}{s}\\
    LES time step&$\Delta t_{\textit{LES}}$&$0.5$ \SI{}{s}\\
    Structural time step&$\Delta t_{\textit{MBS}}$&$0.02$ \SI{}{s}\\
    \hline
         &  \\
         & 
    \end{tabular}
    \label{tab:spec_numerics}
\end{table}

A sub-set of the reference cases has been previously used in an open-loop wind farm power maximization study, \cite{SoodAIAA} and will be used in the next sections to validate the fatigue look-up table for wind turbines in yawed states, and also to showcase the benefits of closed-loop control. These cases have been summarized in Table \ref{tab:reference_cases}, where \textit{PDk} refers to PDBL simulations and \textit{CNk2} and \textit{CNk4} refer to CNBL simulations. The suffix 2 and 4 for the CNBL simulations denotes the strength of the capping inversion. All the reference cases have a surface roughness of $2\times 10^{-4}~m$. The CNBL cases have boundary layer heights of \SI{500}{\metre} and \SI{250}{\metre} for the CNk2 and CNk4 case respectively, and a constant geostrophic wind speed of \SI{12}{\metre\per\second}. Detailed descriptions of the CNBL simulations are available in a previous study.\cite{Sood2020}

\begin{table}
    \caption{Specifications of the reference database}
    \centering
    \begin{tabular}{lcccc}
    \hline
    Case No. & Inflow &Wind direction & Hub height wind speed&Hub height TI\\
    \hline
    1&\textit{PDk}&\SI{0}{\degree}&\SI{9.4}{\metre\per\second}&5.15\%\\
    2&\textit{CNk2}&\SI{300}{\degree}&\SI{11.0}{\metre\per\second}&3.66\%\\
    3&\textit{CNk2}&\SI{330}{\degree}&\SI{11.0}{\metre\per\second}&3.66\%\\
    4&\textit{CNk4}&\SI{0}{\degree}&\SI{11.3}{\metre\per\second}&3.65\%\\
    \hline
         &  \\
         & 
    \end{tabular}
    \label{tab:reference_cases}
\end{table}

\subsection{Model calibration and inflow velocity estimation}

\subsubsection{Calibration}
 
 While analytical wind farm models such as the one defined in Section \ref{wake_model} are very useful for fast analysis and optimization of wind farm operation, they inherently suffer from accuracy issues due to the simplification of the involved physics. Therefore, they require wind farm and site specific parameter calibration to improve their reliability.~\cite{Zhan, Göçmen_2018, Howland2022} In a previous study , the wake expansion parameter $k_w$ from the Bastankhah model was tuned for the TotalControl wind farm database on a per turbine basis, \cite{Sood_tuning} resulting in improved model agreement between the wake model and the reference LES database from Section \ref{ref_database}. However, using this methodology the number of calibration parameters would increase with the size of the wind farm, which could lead to model over-fitting. Additionally, the wake expansion parameter by itself is unable to account for the deflection of the wakes under yawed conditions. Thus in this work, to prevent over-fitting and extend the tuning process to turbines with wake steering as well, we make two refinements to this tuning methodology. First, instead of tuning the wake expansion parameter $k_w$ for all the turbines, we tune a linear relation between wake expansion and turbulence intensity,~\cite{wake_expansion}, i.e.,
 \begin{equation}
     k_w = k_a \cdot I_{rotor} + k_b
 \end{equation}
where $k_a$ and $k_b$ are empirical tuning parameters, and the rotor turbulence intensity is evaluated by the method of Crespo-Hernandez.~\cite{CRESPO199671} Instead of on a per turbine basis, $k_a$ and $k_b$ are tuned for the entire farm. Second, the onset of the far wake region $x_0$ from the Bastankhah model can also be calibrated to improve the wake deflection prediction. According to the model, this distance is given by
\begin{equation}
    \frac{x_0}{D}=\frac{\cos{\gamma}(1+\sqrt{1-C_T})}{\sqrt{2\left[4 \alpha I_{rotor} + 2\beta(1-\sqrt{1-C_T}) \right]}}
\end{equation}
where $\alpha$ and $\beta$ are case specific calibration parameters. Therefore, we can define a calibration vector $\mathbf{\kappa} = [k_a, k_b, \alpha, \beta]$ that can be optimally tuned to minimize the error between the analytical model predictions and the LES power measurements. The minimization problem for parameter tuning can thus be defined as
\begin{equation}
    \min_{{\boldsymbol{\kappa}}} \  \frac{1}{N_t} \sum_{i=1}^{N_t} \left({P_i}^{LES} - {P_i}^{GWM}(\boldsymbol{\kappa})\right)^2 + \lambda \sum_{i=1}^{N_t} {\boldsymbol{\kappa}_i}^2
\end{equation}
To prevent over-fitting of the model, a regularizing penalty term is introduced in the above optimization problem using ridge-regression through the ridge parameter $\lambda$.~\cite{ridge, Sood_tuning} In this work, we utilize the same value of the ridge parameter as a previous study for parameter tuning of the TC RWP,~\cite{Sood_tuning} i.e. $\lambda = 2$, as it still lead to a reduction in power errors after calibration when tuning the new model parameters defined in $\boldsymbol{\kappa}$. Though it must be noted that ridge regression is typically useful when all the tuning parameters have a similar magnitude, while this is not the case currently. In practice, a more robust method such as Bayesian inference could be used to obtain the optimal value of $\lambda$,~\cite{bayesian} but this is not investigated in the current work. The minimization problem is subsequently solved using the SLSQP solver from the SciPy package. \cite{2020SciPy-NMeth}

\subsubsection{Estimation}

In order to optimize the performance of the wind farm, the state model defined in Section \ref{wake_model}, requires information regarding the current operating conditions of the turbines within the farm. This includes the following:
\begin{itemize}
    \item Power production of all turbines
    \item Current control states of all turbines (yaw angle, pitch angle, rotational speed)
    \item Inflow wind speed
    \item Inflow turbulent intensity
    \item Wind direction.
\end{itemize}
While the first 2 items are readily available as measurements or outputs of the wind farm, accurate measurements of the latter 3 items are not usually available in commercial wind farms and need to be estimated. In the current work, we focus on estimating the inflow wind speed of the farm, while the turbulence intensity and wind direction are assumed to remain statistically constant during the simulation run and taken directly from the reference database detailed in Table \ref{tab:reference_cases}. The inflow wind speed is estimated by the following expression,
\begin{equation}
    U_b = argmin_{\tilde{U_b}} \left(\frac{1}{N_U}\sum\left(\bar{P_i} - \hat{P_i}(\phi,\widetilde{U_b},\bar{\gamma_i}) \right)^2 \right)
\end{equation}
where $\bar{P_i}$ is the measured power output from the virtual LES environment, averaged of the sampling time, and $\hat{P_i}$ is the predicted power by the wind farm state model, per turbine $i$. $N_U$ is a set of the upstream wind turbines of the farm, which has been previously determined based on the wind direction $\phi$. The averaging time for the measurements depends upon the sampling time $t_s$, which is the regular constant time interval after which the closed-loop control framework is executed.

\subsection{Observablity}

Due to the simplifications made in the framework, it is important to remark upon the issue of observability. In real world implementation of the closed-loop framework, sufficient information needs to be measurable to estimate the ambient conditions and model parameters. In the current framework (see Figure \ref{fig:closed_loop_formulation}), we first estimate the free stream velocity based solely on the power production of the upstream, un-waked turbines, and then subsequently estimate the model parameters based on the power error for all the turbines within the farm. This is an idealized situation, since the model is provided information  a priori regarding the wind direction, the resulting set of upstream turbines and the turbulence intensity. Additionally, the inflow velocities from the selected inflow database result in only a region 2 operation of the DTU turbines, therefore errors in estimation of the wind speed due to region 3 operation are also not a concern.  Observability issues would indeed need to be addressed in future work if these idealized assumptions are not followed. The readers are directed to the work of Doekemeijer for more information regarding the issue of observability in closed-loop control~\cite{observability}.
 
 \section{Fatigue Look-Up Table}{\label{fatigue_lut_section}}
 
\begin{figure}
	\centering
	\includegraphics[width=4in]{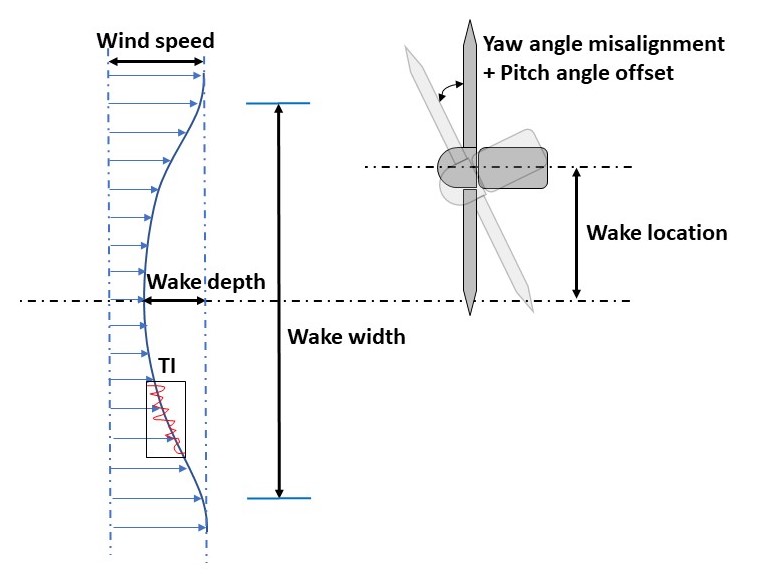}
	\caption[Scheme of the input parameters of the load lookup table.]{Scheme of the input parameters of the load lookup table based on a study by TU Delft. \cite{mendez2019validation} The input parameters are wind speed, turbulence intensity (TI), yaw angle misalignment, pitch angle offset, wake deficit depth, wake width and wake center location.}
	\label{fig:loadLUTfigure}
\end{figure}

  To determine the fatigue in different turbine components, a look-up table needs to be developed which covers the different operational states that a wind turbine operating within an offshore farm environment experiences. The LUT needs to be valid across a range of values for different turbine yaw and pitch angles, while operating under different inflow conditions. Fatigue loads are dependent on several parameters such as wind speed, turbulence intensity, yaw angle misalignment, pitch angle offset, wake characteristics, wind shear, wind veering, etc. Ideally, the load lookup table should contain fatigue loads for any combination of values for these parameters, however acquiring such a detailed lookup table is not feasible. Therefore, a discrete set of values is chosen for every input parameter. Simulations to develop the load LUT are carried out using the OpenFAST open-source wind turbine simulation tool developed by NREL.~\cite{OpenFAST2021} As the structural response is dependent on the turbine material properties and design, OpenFAST setup files of the DTU \SI{10}{\mega\watt} RWT are used for the simulations to match the turbine used in the reference LES cases. \cite{DTU10MWRWTrepository} The turbine operation and controller tuning is carried out using the ROSCO controller and toolbox.~\cite{mulders2018delft,ROSCO_toolbox_2021} In this way, a 7D fatigue load lookup table is created whose input dimensions are based on a study by TU Delft~\cite{mendez2019validation} and are visualized in Figure ~\ref{fig:loadLUTfigure}. For simplicity, variations in parameters such as wind shear, wind veering, turbulence length scales and time scales are not considered in this table. The inclusion of these parameters may be of interest in future research, as they can have a significant effect on the turbine loads and fatigue.
  
\begin{figure}
	\centering
	\includegraphics[width=6in]{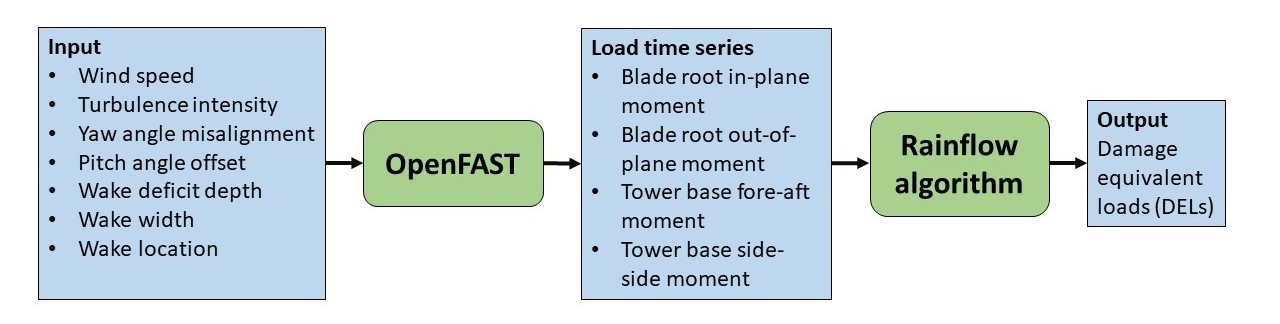}
	\caption[Scheme of the design of a fatigue load lookup table.]{Scheme of the design of a fatigue load lookup table with seven input dimensions and four output dimensions (four DELs for each set of input parameter values, one load time series corresponds to one DEL).}
	\label{fig:loadLUTinout}
\end{figure}

  The output channels of the LUT table are selected to be the mean and fatigue loads associated with the blade root and tower base. While it is possible to obtain more load channels from OpenFAST, we restrict our analysis to these turbine components for simplicity. OpenFAST is used to generate 10 minute load time series, and a Rainflow counting algorithm~\cite{rainflowMatlab} is used to obtain the DEL for each load time series.~\cite{freebury2000determining} A schematic showing the procedure of developing the LUT and it's different parameters is shown in Figure~\ref{fig:loadLUTinout}.
  
\subsection{Range of the input parameters}

\begin{figure}
	\centering
	\subfloat{\includegraphics[width=0.40\textwidth]{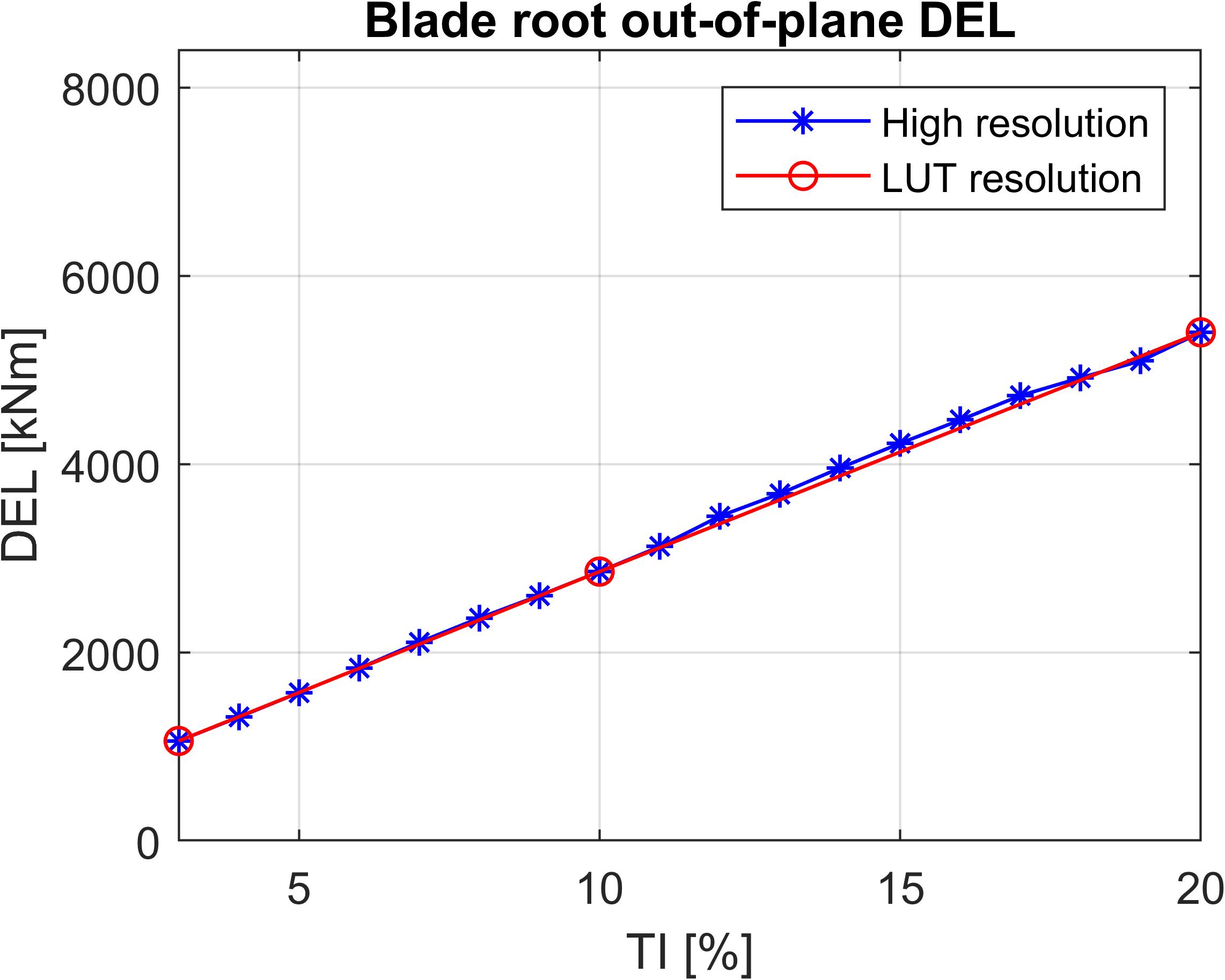}}
        \hspace{10pt}
	\subfloat{\includegraphics[width=0.40\textwidth]{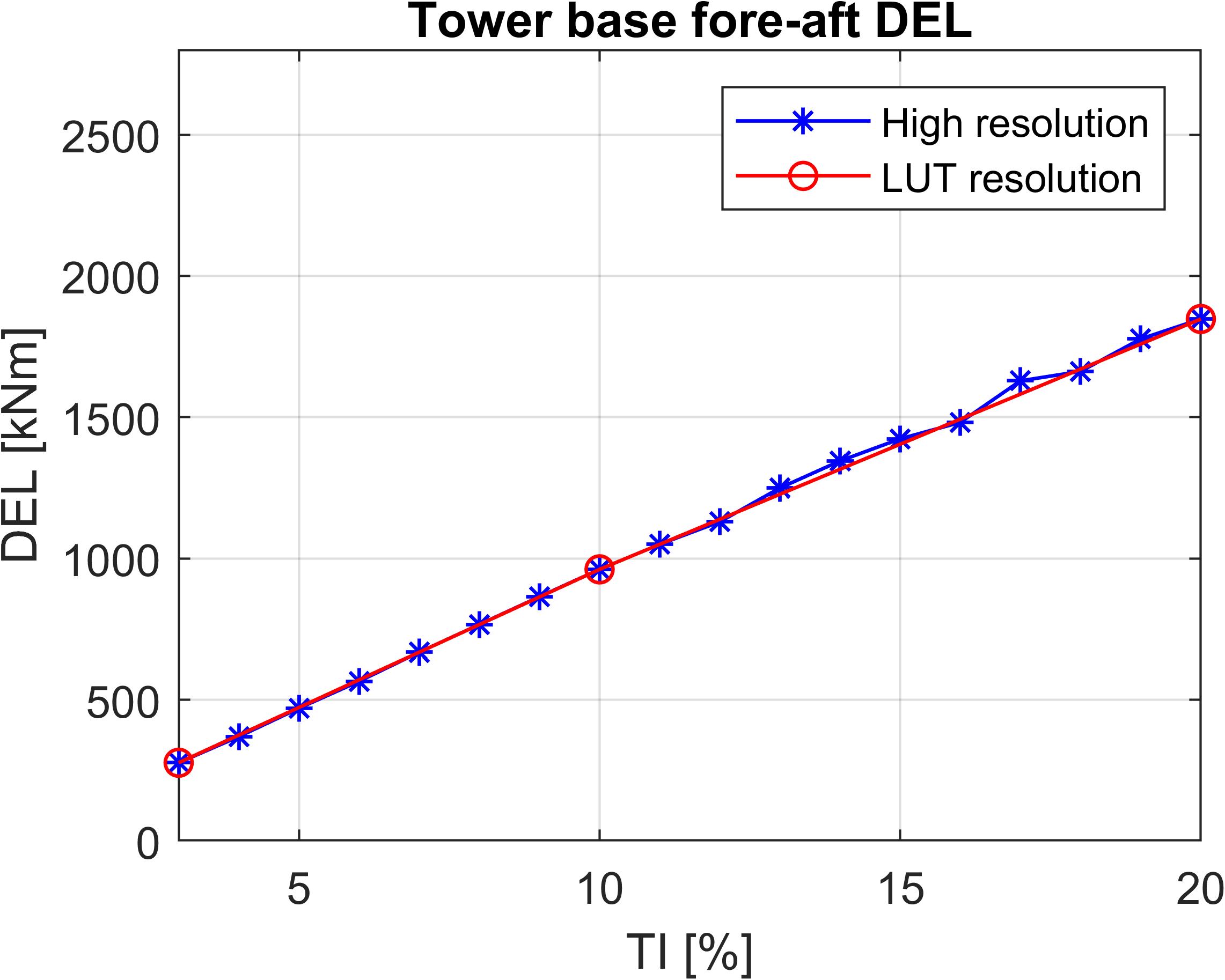}\label{fig:RootMOoPb_DEL_wsp}}
        \hspace{10pt}
	\caption[Test]{Damage equivalent load as a function of turbulence intensity. Left: blade root out of plane DEL, right: tower base fore-aft DEL.}
	\label{fig:RootMOoP_DEL_ti}
\end{figure}

This section details the input parameters used to create the LUT and their ranges. In total, seven dimensions are included: wind speed, turbulence intensity, yaw angle misalignment, pitch angle offset, wake deficit depth, wake width and wake center location. To determine the range and resolution of the other dimensions, a sensitivity analysis is conducted for each parameter in which each parameter is varied in small steps, while keeping the other dimensions constant. For each dimension, a fine resolution simulation is taken as the reference, and then a coarser resolution of each parameter which can accurately represent the trends of the fatigue loads is determined. This resolution reduction is necessary to reduce the number of simulations required to develop the LUT. The extreme values of each dimension are determined based on observations made from the reference LES cases given in Table~\ref{tab:reference_cases} to avoid extrapolation.  In this section, the results of turbulence intensity, wind speed and yaw angle sensitivity analysis are presented, while the results for the other dimensions are provided in Appendix ~\ref{app1}. 

The relation between damage equivalent loads and turbulence intensity ($I$) is very linear. This can be seen in Figure~\ref{fig:RootMOoP_DEL_ti}, which shows the blade root out-of-plane DEL as a function of turbulence intensity. This trend is also shared by the blade root in-plane and tower base DELs. Three values for the turbulence intensity are retained to represent the linear DEL trend, namely 3, 10 and 20\%. In addition to the choice of turbulence intensity values, it is important to specify the number of turbulence seeds for every turbulence intensity and/or the length of the simulation time in OpenFAST and TurbSim. The IEC standards recommend using at least six different turbulence seeds and a 10 minute simulation time length.~\cite{IEC61400-1,IEC61400-3} However, a study by NREL shows that this amount of turbulence seeds is not always sufficient.~\cite{haid2013simulation} Hence, a separate analysis is made to choose a suitable amount of turbulence seeds, 
based on the relation between the simulation time $T$ and the turbulence intensity $I$.~\cite{Pope_2001} The latter is defined as:
\begin{equation}\label{eq:ti}
	I = \frac{\langle u^2 \rangle^{1/2}}{\langle U \rangle}
\end{equation}
with $U$ the turbulent wind velocity, $\langle U \rangle$ the average velocity and $u = U - \langle U \rangle$ the velocity fluctuation. Further, a similarity for the variance of the time average $\sigma^2_{\langle U \rangle, T}$ can be found using $\mbox{var}(U) = \langle U^2 \rangle - \langle U \rangle^2 = \langle u^2 \rangle$,\cite{Pope_2001}
\begin{equation}\label{eq:var_time_avg}
	\sigma^2_{\langle U \rangle, T} \sim \frac{\langle u^2 \rangle 2 \bar{\tau}}{T}
\end{equation}
with $\bar{\tau}$ the integral timescale and $T$ the length of the simulation time. For $T \rightarrow \infty$ the time averages converge. A limit can be defined on the standard deviation $\sigma_{\langle U \rangle, T}$ as follows,
\begin{equation}\label{eq:standard_deviation}
	\sigma_{\langle U \rangle, T} = \epsilon \langle U \rangle
\end{equation}
with $\epsilon$ a small error value. Equations \ref{eq:ti} - \ref{eq:standard_deviation} can be combined into one equation:
\begin{equation}\label{eq:T_ti_relation}
	T \sim I^2 \frac{2\bar{\tau}}{\epsilon^2}
\end{equation}
Equation~\ref{eq:T_ti_relation} shows that the simulation time length depends on the turbulence intensity squared. Under the hypothesis of ergodic turbulence, the time average can be replaced by a sample average. Increasing the simulation time length thus comes down to increasing the amount of simulations with different turbulence seeds. Based on a convergence study shown in Appendix~\ref{app2}, for this work three seeds are used for $I = 3\%$, six seeds for $I = 10\%$ and twelve seeds for $I = 20\%$. 

\begin{table}
	\centering
	\caption[Values of load lookup table input parameters.]{Values of load lookup table input parameters.}
	\label{tab:LUT_values}
	\begin{tabular}{@{}llc@{}} \toprule
		\textbf{Input parameter} & \textbf{Values} & \textbf{Units} \\ \midrule
		Wind speed & $\left[4, 5, 6, 7, 8, 9, 10, 11, 12, 13, 19, 25\right]$ & m/s \\
		Turbulence intensity & $\left[3, 10, 20\right]$ & \% \\
		Yaw angle misalignment & $\left[-30, -20, -10, 0, 10, 20, 30\right]$ & degrees \\
		Pitch angle offset & $\left[-6, -4, -2, 0, 2, 4, 6\right]$ & degrees \\
		Wake deficit depth & $\left[0, 0.3, 0.5\right]$ & - \\
		Wake width & $\left[0.65, 1.2, 1.73\right]$ & - \\
		Wake center location & $\left[-1.5, -0.6, 0, 0.6, 1.5\right]$ & D (diameters)\\
		\bottomrule
	\end{tabular}
\end{table}

\begin{figure}
	\centering
	\subfloat{\includegraphics[width=0.32\textwidth]{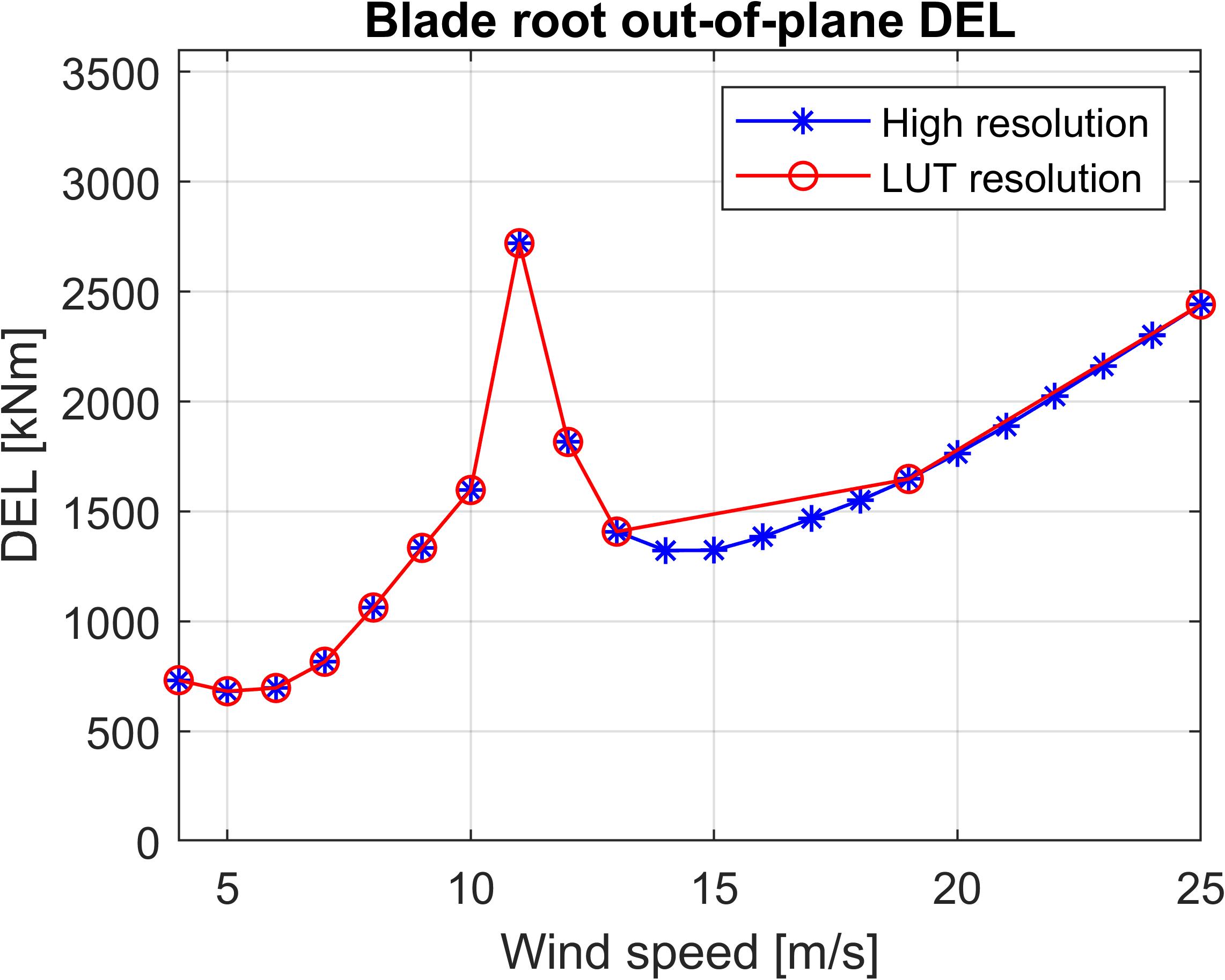}\label{fig:RootMOoP_DEL_wsp}}
        \hspace{5pt}
	\subfloat{\includegraphics[width=0.32\textwidth]{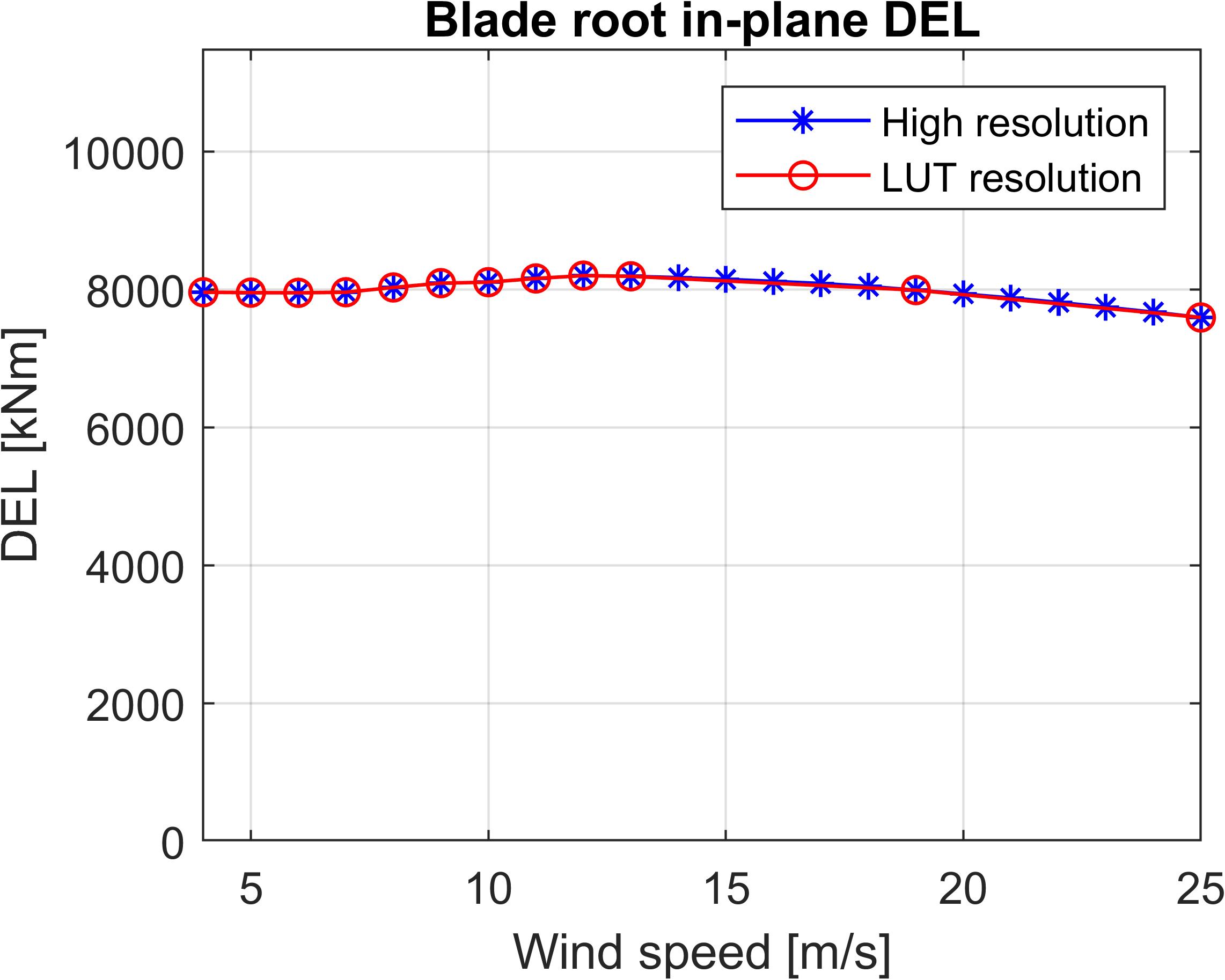}}
        \hspace{5pt}
	\subfloat{\includegraphics[width=0.32\textwidth]{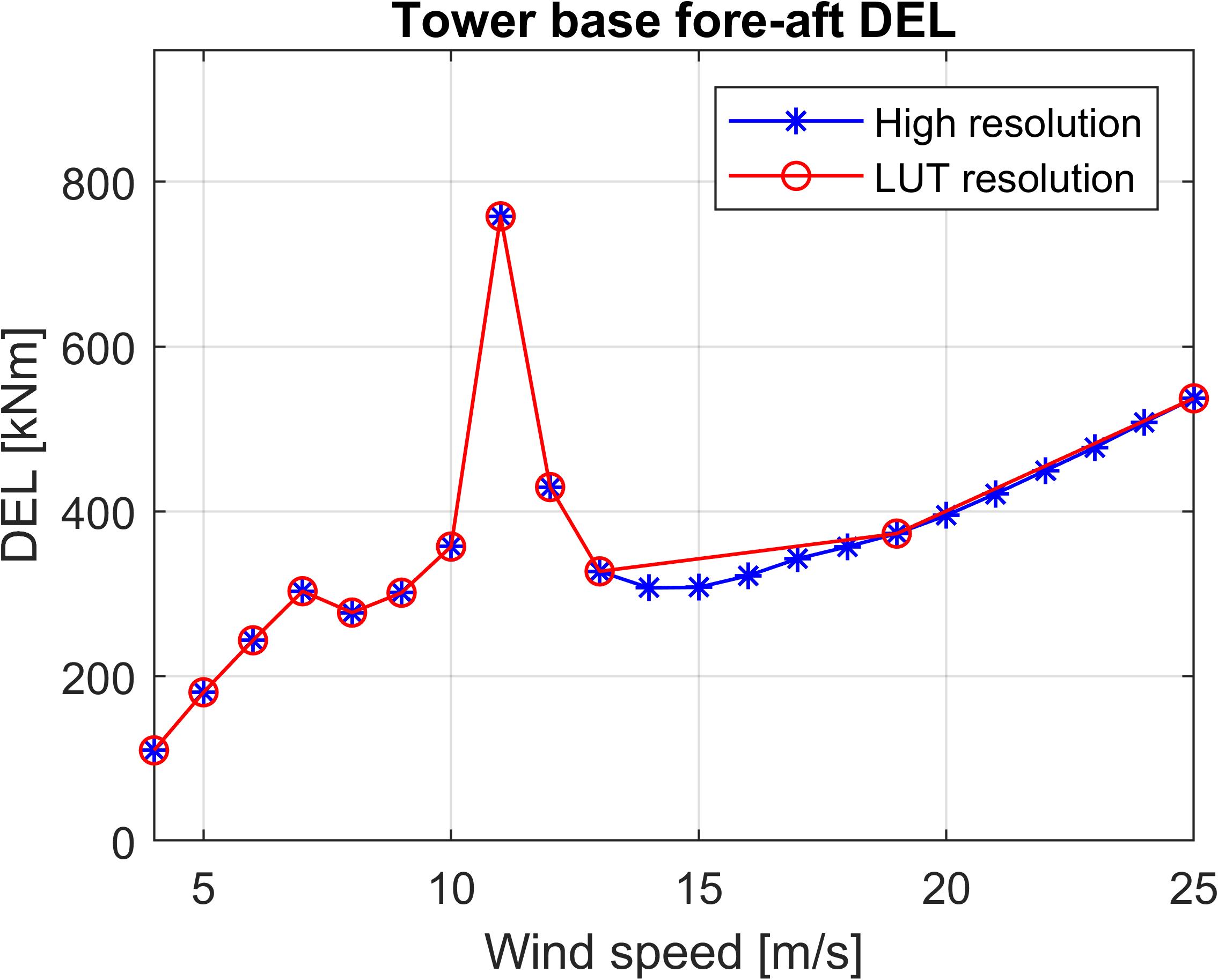}\label{fig:TwrBsSSb_DEL_wsp}}
    \\
	\subfloat{\includegraphics[width=0.32\textwidth]{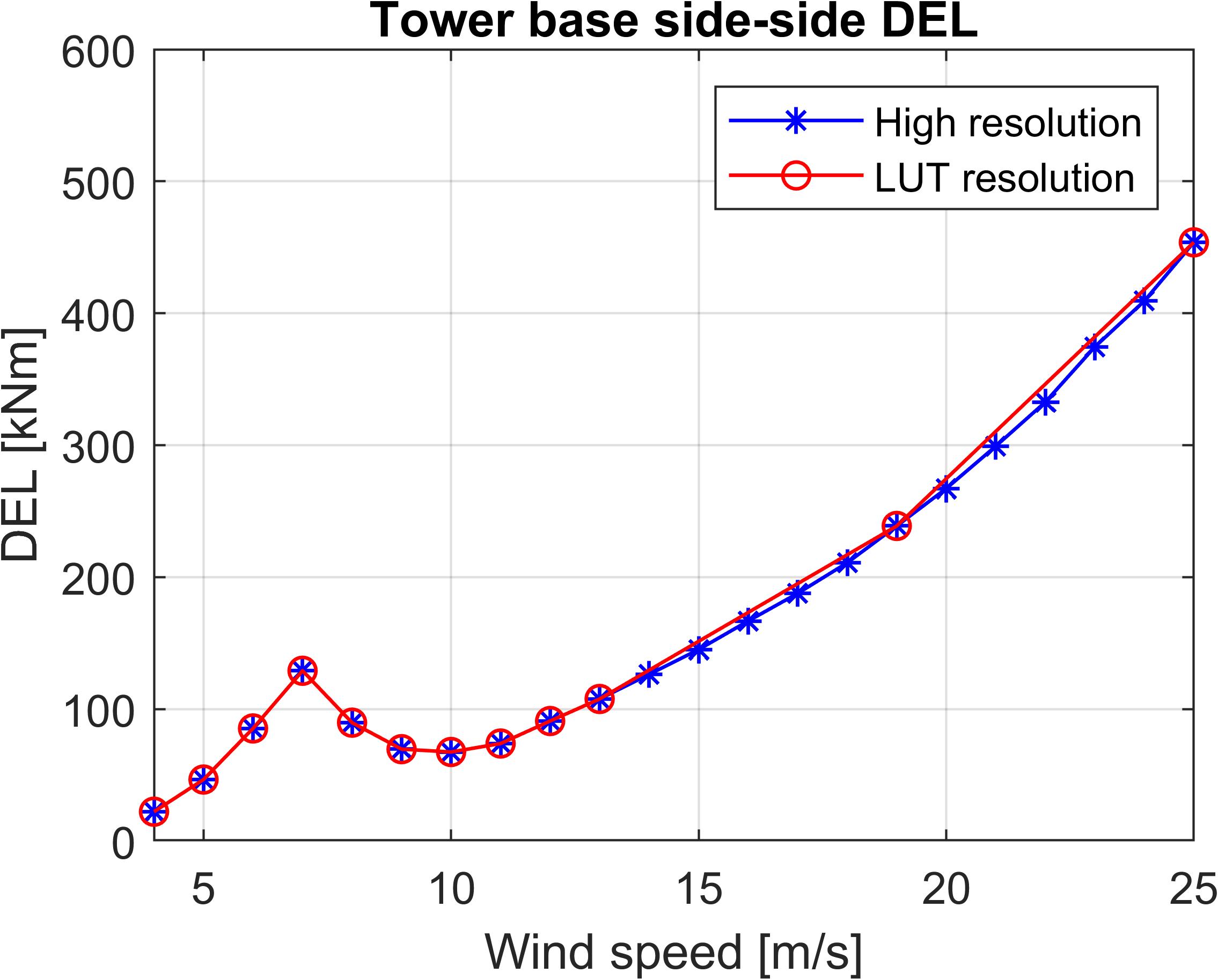}\label{fig:TwrBsSS_DEL_wsp}}
        \hspace{5pt}
 	\subfloat{\includegraphics[width=0.32\textwidth]{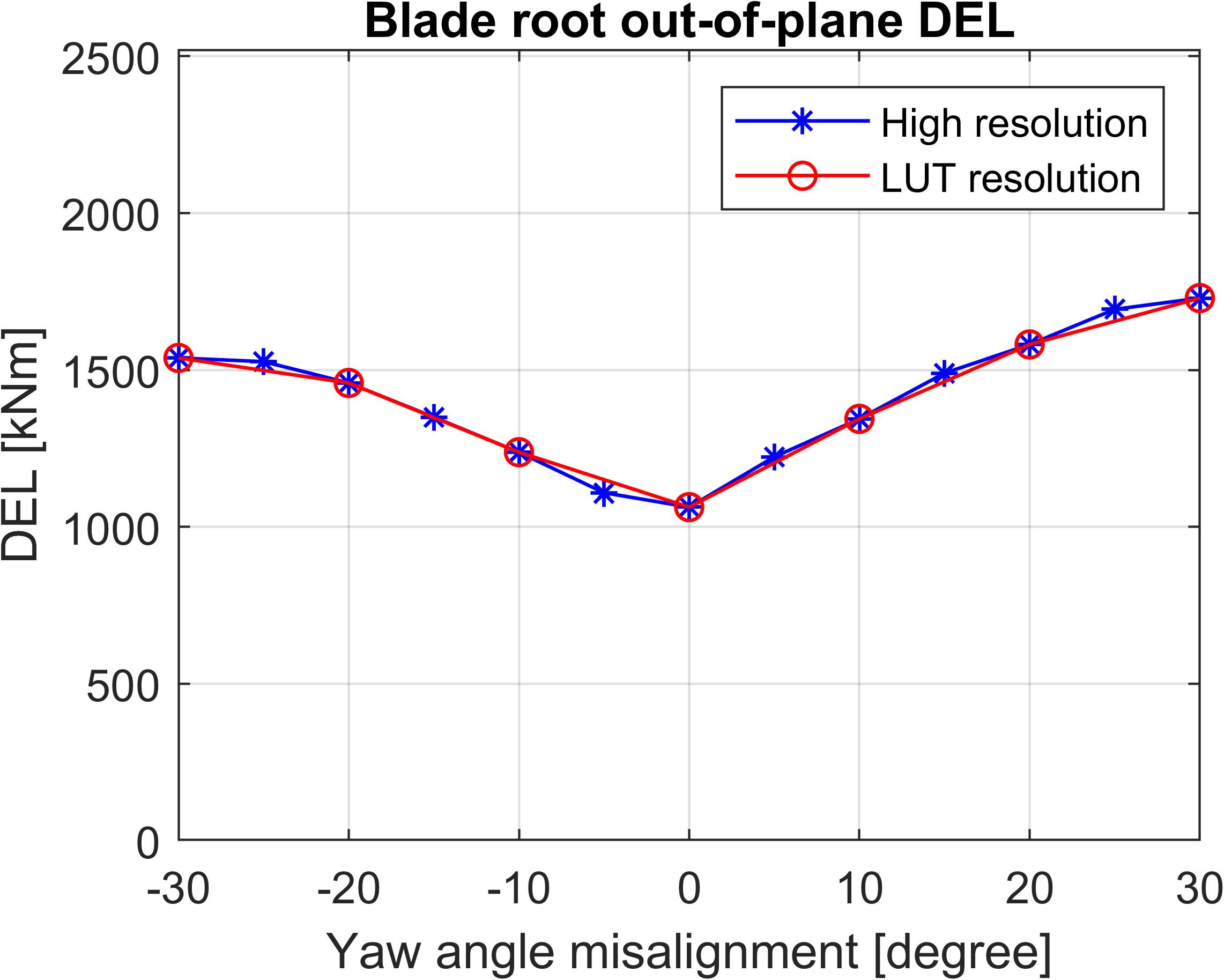}\label{fig:RootMOoP_DEL_ya}}
        \hspace{5pt}
	\caption[]{(a-b) Blade root and (c-d) tower base DEL as a function of wind speed. (e) Blade root out-of-plane damage equivalent load as a function of yaw angle misalignment.}
	\label{fig:DEL_wsp}
\end{figure}

The extreme values for the wind speed range are based on the cut-in and cut-out wind speeds of the DTU \SI{10}{\mega\watt} RWT, hence are set to 4 and 25 \SI{}{\metre\per\second}. Between this range, a 1 \SI{}{\metre\per\second} resolution is used until 13 \SI{}{\metre\per\second} as it just beyond the rated wind speed of the turbine, beyond which the trend was observed to become linear, therefore only 2 additional data points are used. The resulting fatigue load trends are shown in Figure~\ref{fig:DEL_wsp}. Figures \ref{fig:RootMOoP_DEL_wsp} - \ref{fig:TwrBsSS_DEL_wsp} show a comparison between the high resolution DELs and the linearly interpolated LUT resolution DELs. For yaw angle variation, the blade root out-of-plane DEL trend depicted in Figure \ref{fig:RootMOoP_DEL_ya} has a curved V shape. Since the yaw angle misalignment trend is not linear, we find that a minimum resolution of 10 degrees was required to capture the trends across the operation range.

Based on the resolutions chosen across the 7 dimensions, a total of 555,660 simulations would be needed if every combination of input parameters would need an aeroelastic simulation in OpenFast. However, when the wake deficit depth equals zero, no wake is present and the parameters wake width and wake center location become irrelevant. Consequently, 382,788 simulations are required, which are performed on the Tier-2 cluster Genius from the Flemish Supercomputer Center (VSC). Finally, a multidimensional linear interpolation scheme is used to interpolate the DELs stored in the lookup table, without the use of extrapolation.

\subsection{Validation of load LUT}

\begin{figure}
    \centering
    \includegraphics[width=0.8\textwidth]{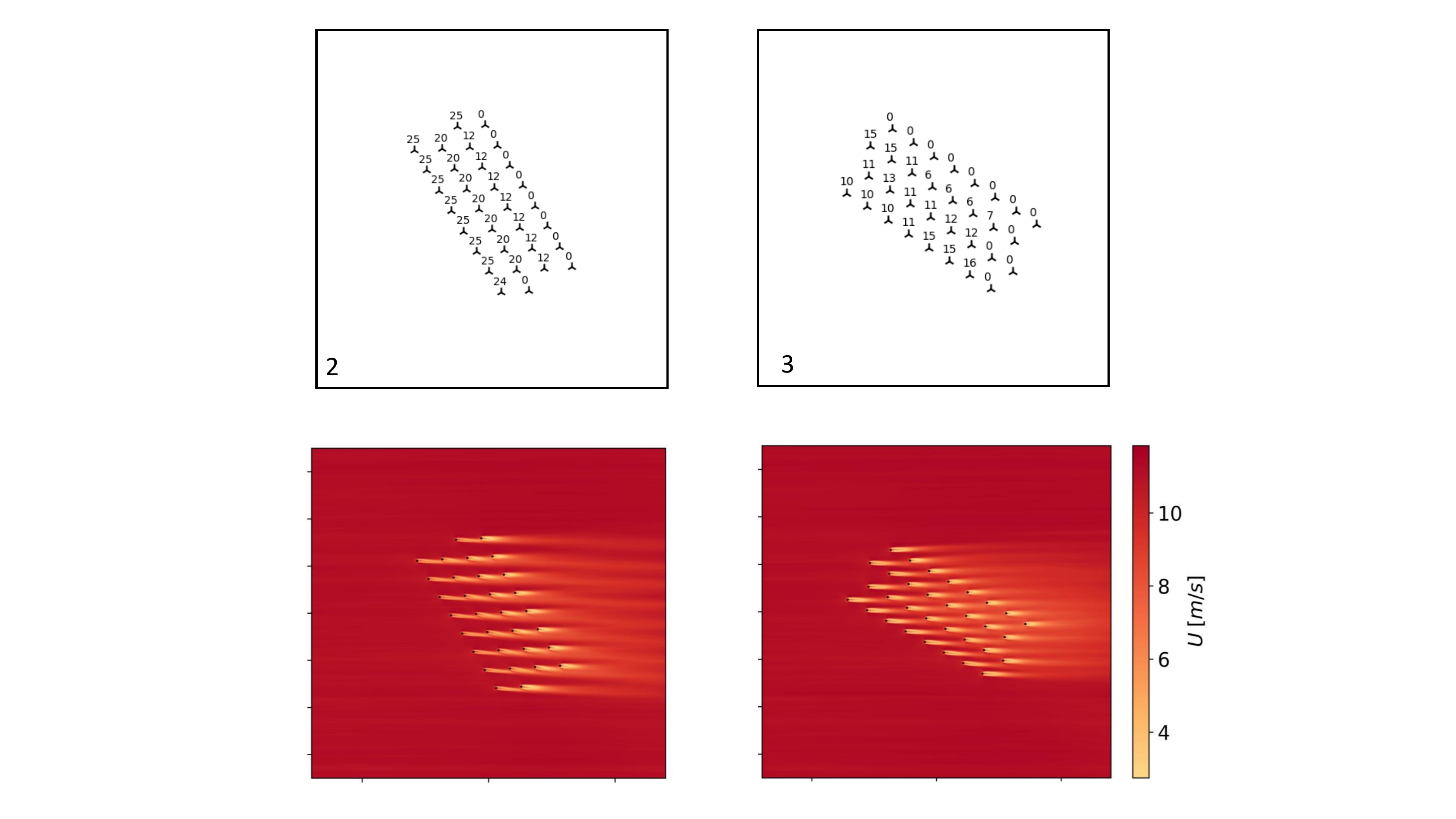}
    \caption{Optimal wake steering set points (top) and the resulting LES time averaged flow fields (bottom) for cases 2 and 3 from Table \ref{ref_database}, obtained using a previous open-loop study.\cite{SoodAIAA}}
    \label{fig:cnk2_yaw_set_points}
\end{figure}

\begin{figure}
    \centering
    \includegraphics[trim=0 150 0 120,clip,width=\textwidth]{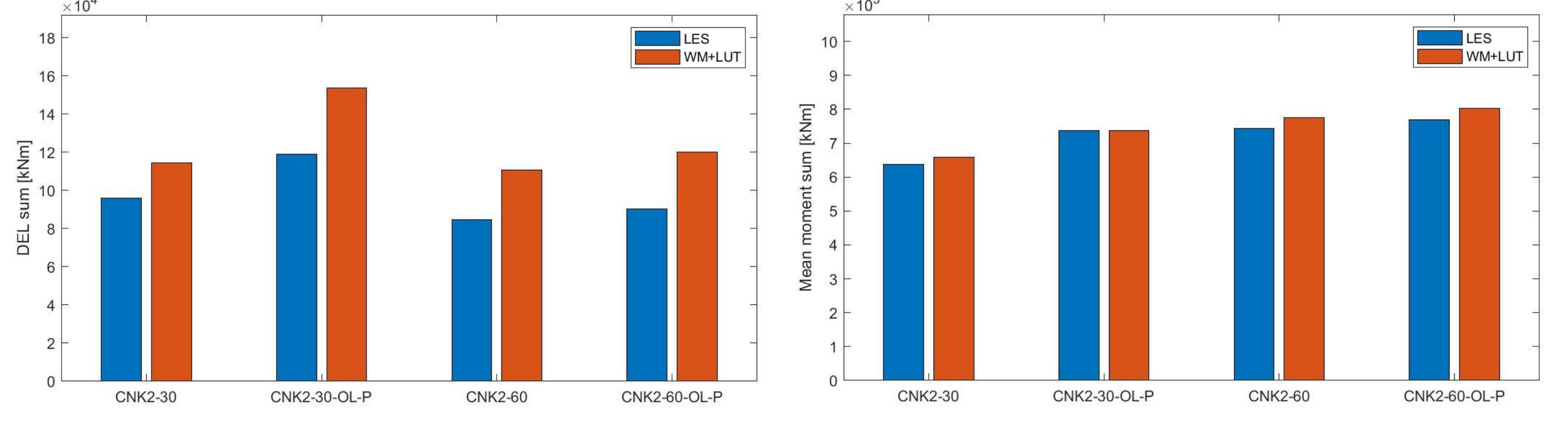}
    \caption{Comparison of total blade root out plane fatigue and mean moments between the LES and the wake model with load LUT.}
    \label{fig:LUT_total_increase}
\end{figure}

\begin{figure}
    \centering
    \includegraphics[trim=150 0 150 0,clip,width=0.8\textwidth]{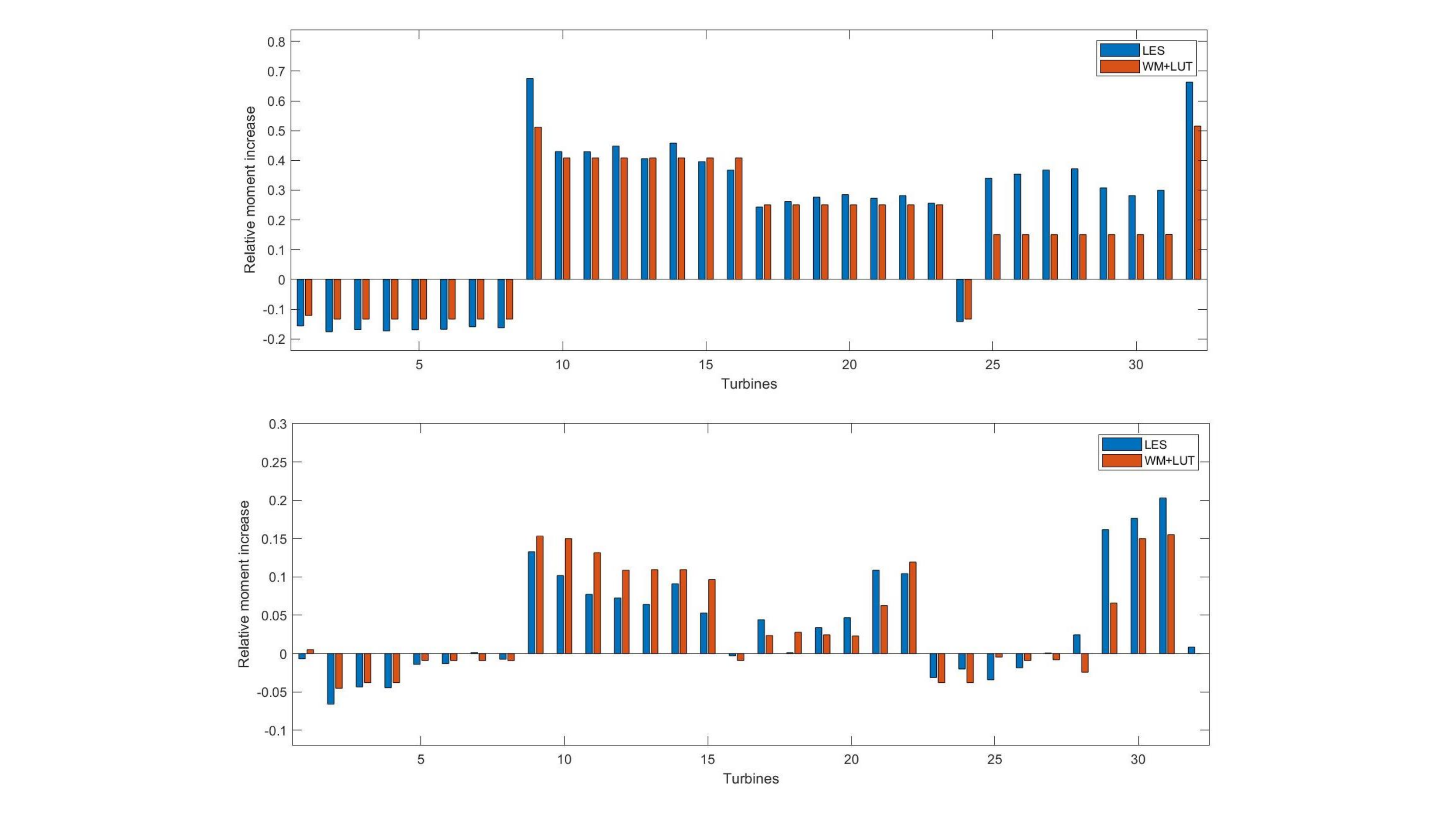}
    \caption{Comparison of individual turbine relative blade root out of plane moment increase for CNk2-30 (top) and CNk2-60 (bottom), with and without wake steering.}
    \label{fig:LUT_indvl_increase}
\end{figure}

To validate the load LUT, the results are compared against the reference LES cases defined in Section 2.3, in which simulations are performed using the TCRWP (see Figure \ref{fig:tc_rwp}). The CNk2 cases, i.e cases 2 and 3 from Table \ref{tab:reference_cases} are used as the reference normal operation reference. Their corresponding optimal wake steering open-loop wake steering results, the yaw set point distribution and time averaged flow fields, which are shown in Figure \ref{fig:cnk2_yaw_set_points}, are also used to compare the performance of the wake model plus LUT for coordinated wake steering cases. A comparison of the total blade root flapwise fatigue and mean load increase is shown in Figure \ref{fig:LUT_total_increase}, while a per turbine breakdown is shown in Figure \ref{fig:LUT_indvl_increase}. The open-loop power optimization results are designated by including the letters 'OL-P' after the case name, which is written as the inflow followed by the angle of farm rotation to achieve the desired wind direction. From the results, it can be seen that while slight over prediction is observed in the total fatigue loads with upto 10\% over prediction , the mean moments match well for both the greedy and the open-loop wake steering cases with below 5\% error. On an individual turbine level, while errors greater than 100\% are present for some turbines, the overall trend of load increase or decrease across the farm is captured well. The difference in fatigue loading could be attributed to the different controllers used for developing the LUT and performing the reference wind farm simulations, which could cause difference in the moment variances and thus fatigue. Thus while not extremely accurate, it will be established in the following section that the developed load LUT enables combined power and load optimizations and leads to reduction in turbine loading when incorporated into the closed-loop control procedure.

\section{Wind farm performance optimization}

\begin{figure}
    \centering
    \includegraphics[width=0.7\textwidth]{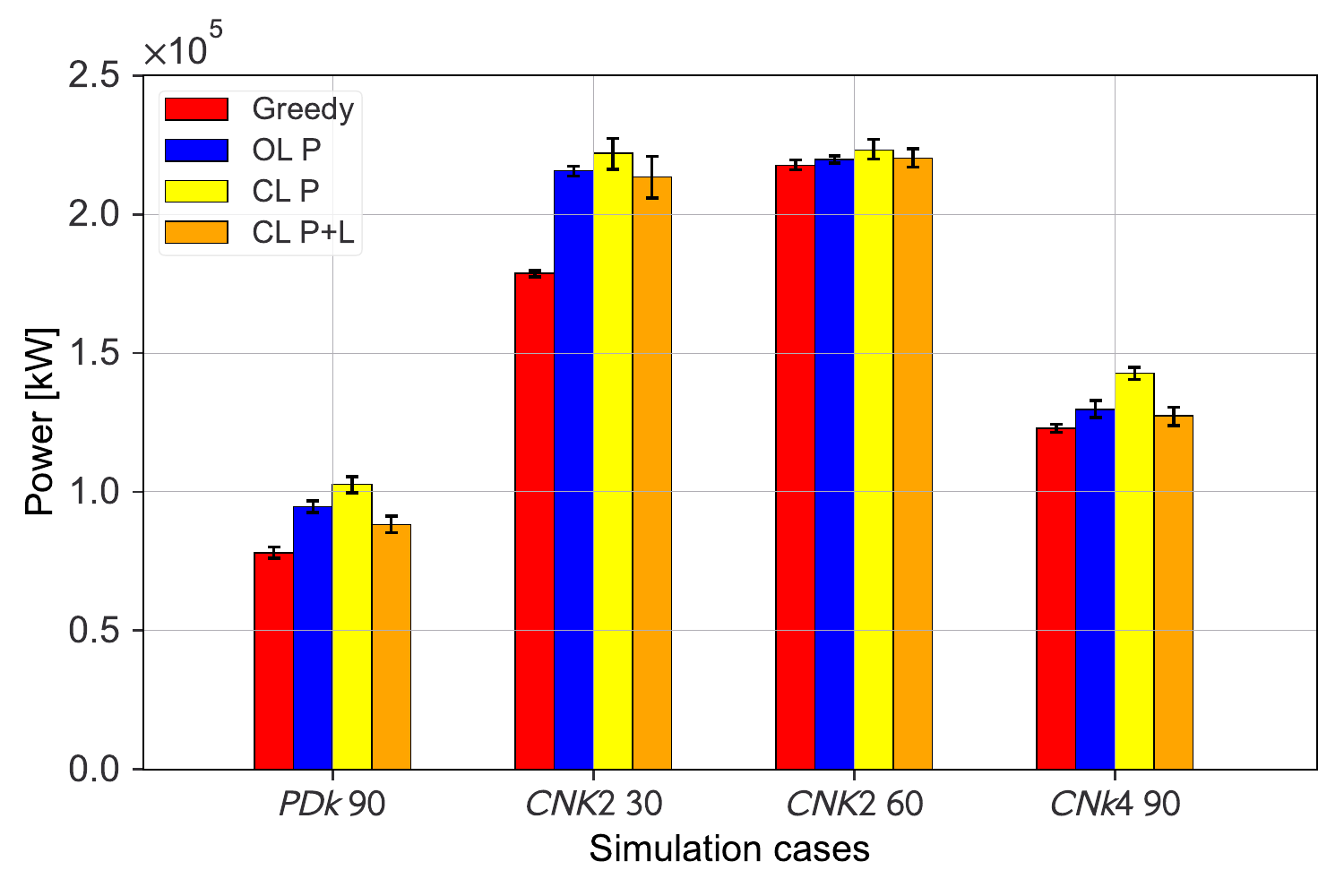}
    \caption[Comparison of total wind farm power gains through closed-loop control (CL) against greedy operation and open-loop (OL) control.]{Comparison of total wind farm power gains through closed-loop control (CL) against greedy operation and open-loop (OL) control. Optimized cases are for either power optimization (P) or combined power and loads optimization (P+L). Error bars represent 95\% confidence interval, obtained using the block bootstrap method for power time series~\cite{sood_lillgrund}.}
    \label{fig:total_power_gains}
\end{figure}

Total wind farm power gains across the 16 cases after 1 hour of wind farm operation are shown in Figure~\ref{fig:total_power_gains}. It can be seen that amongst all the cases, the $30^\circ$ farm orientation has the highest potential for power gains, about 26\%, and lowest for the $60^\circ$ degree orientation, about 5\%. This is due to the difference in the number of turbines operating in a waked state across the two orientations (see Figure~\ref{fig:cnk2_yaw_set_points}), which leads to a difference in potential for power gains through wake steering. Across all the cases, it can be seen that the closed-loop controller out performs both the greedy operation and the open-loop controller. The main contributor to this effect, is the calibration of the wake model parameters which improves it's capability for predicting and optimizing the performance of the farm~\cite{Howland2022, Sood_tuning}. The highest relative gains can be seen for $90^\circ$ farm orientation, which is characterized by the deepest array of aligned turbines of all the cases, which is a situation in which wake models are known to suffer performance issues~\cite{deep_array}. 

\begin{figure}
    \centering
    \includegraphics[width=0.9\textwidth]{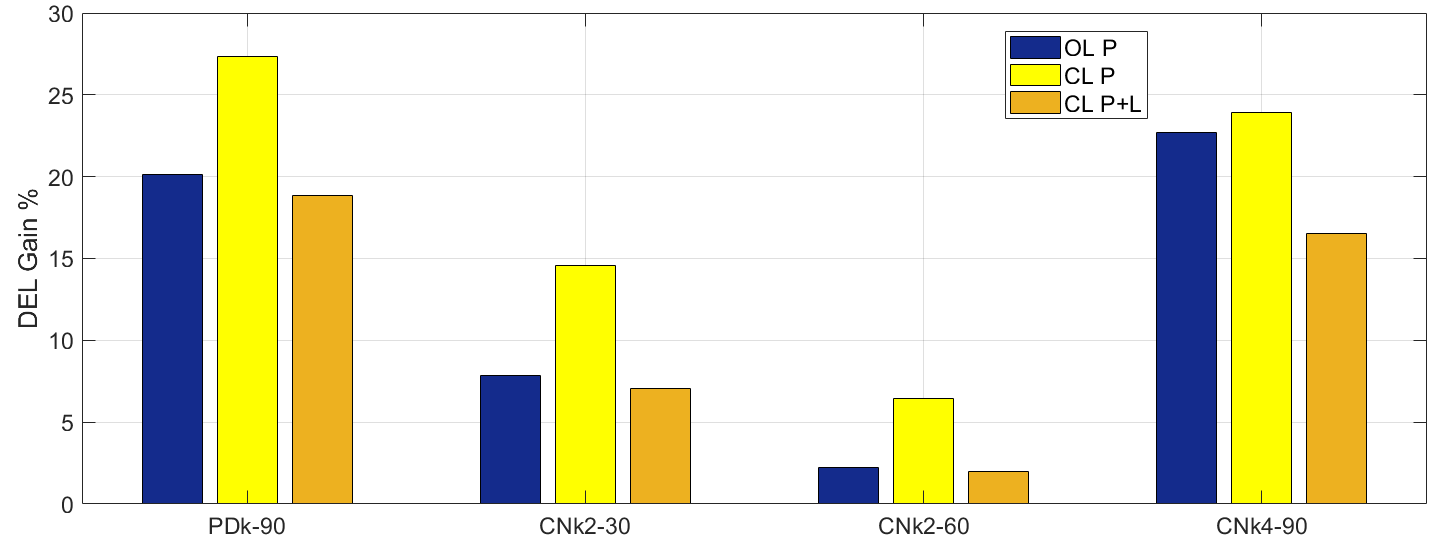}
    \caption{Effect of including loads in the cost function on total farm blade root flap wise fatigue gains.}
    \label{fig:total_fatigue_gains}
\end{figure}

The cumulative blade root fatigue damage across the wind farm for all the cases is evaluated and presented in Figure~\ref{fig:total_fatigue_gains}. The open loop operation have static yaw angles, while the yaw angles in the closed-loop cases are dynamically changing. Changing yaw angles lead to a higher variation in the moments across the turbine blades, resulting in the higher DEL values observed for the closed-loop power optimization cases. As the closed-loop scenarios do lead to converged yaw angles towards the end of the simulation runs, this effect may be short lived and may not be significant for longer simulation times, depending on how fast the ambient conditions change in the field. It can also be observed that the DEL increases are the highest for the cases CNk4 and PDk-90. This is due to the fully aligned wind farm configurations for these two cases which leads to the deepest turbine arrays, thus a potential for the largest number of upstream turbines operating in a yawed state, and also downstream turbines operating in a partially waked state. Both these situations lead to an increase in turbine fatigue due to cyclic variations in moments and hence an increase in DEL. This may not be the case if the turbines were allowed to yaw more than 30 degrees which would reduce the partial yaw operation for downstream turbines, but yaw angles greater than 30 degrees are generally considered too detrimental for the upstream turbines and thus avoided.

Additionally, while the combined power and load optimization leads to a reduction in farm power production when compared to pure closed-loop power production gains, the power gains are still comparable to the open-loop gains. It can be observed that the cumulative blade root fatigue across the farm decreases or remains similar to the open-loop case, while it is lower for all cases when compared to pure closed-loop power optimization. Thus by including fatigue loading in the cost function, while the obtained power gains are lower than closed-loop power optimization, the gains are still comparable to open-loop gains while sustaining lower fatigue damage.

\begin{figure}
    \centering
    \includegraphics[trim=0 100 0 100,clip,width=\textwidth]{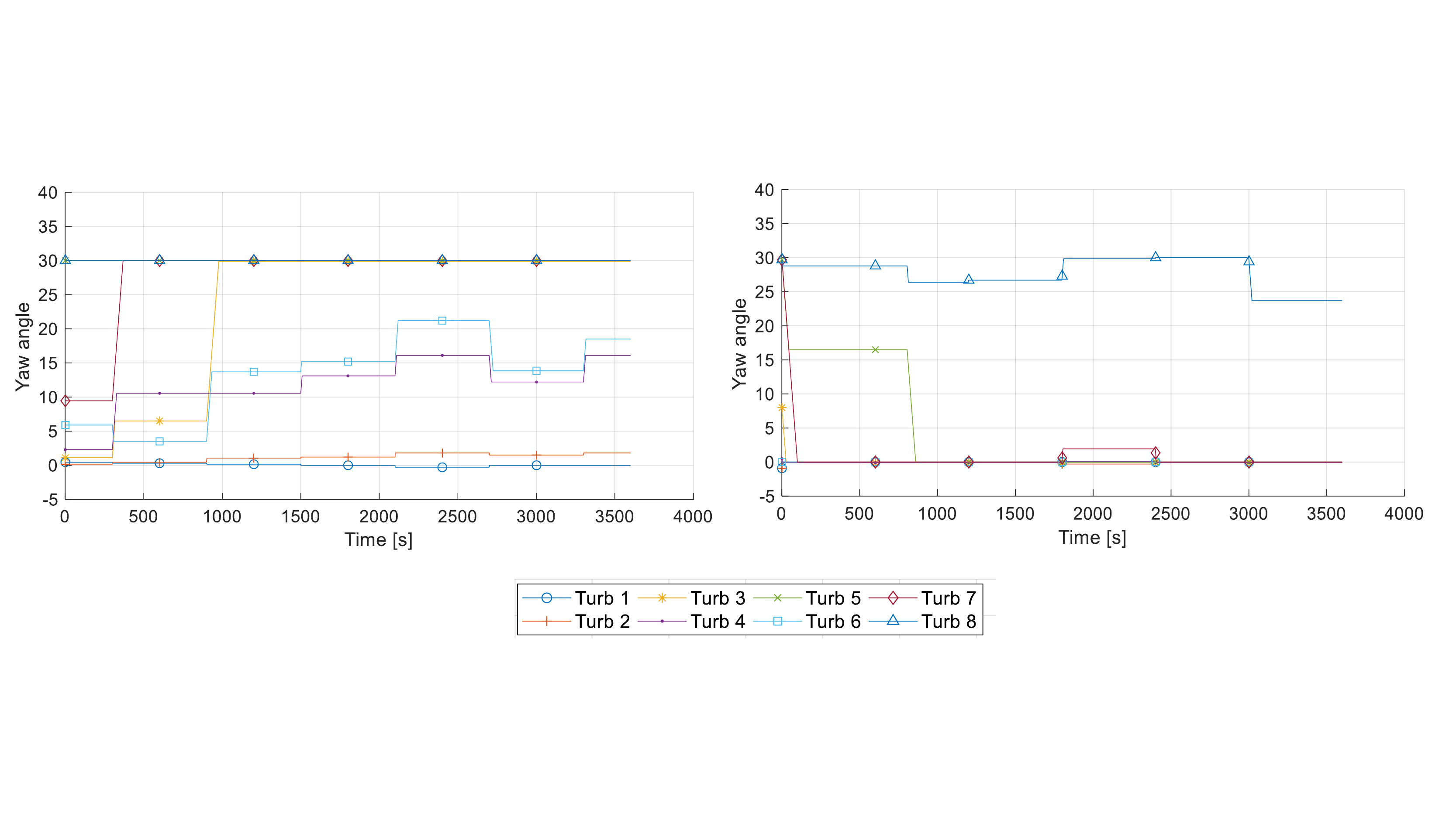}
    \caption[Evolution of yaw angles across one row of turbines.]{Evolution of yaw angles across one row of turbines for the PDk-90-CL-P (left) and PDK-90-CL-P+L (right). Turbine 8 is the most upstream turbine.}
    \label{fig:yaw_time_series}
\end{figure}

The dynamics of the closed-loop framework are further analyzed by evaluating the evolution of the wake steering angles and the calibration parameters. Change in yaw angles for the PDk 90 case as a result of the closed-loop control framework are shown in Figure~\ref{fig:yaw_time_series}. It can be observed that the yaw angles achieve statistically stable values by the end of the simulation time period. For yaw combined power and loads optimization, it is interesting to note that only the most upstream turbine performs wake steering while all the downstream turbines reach a zero yaw state. The resulting wake steering distribution leads to reduced power gains when compared to the pure power maximization case, but the gains are still comparable to the open-loop simulation. However, it must be noted that the optimal yaw set points and the resulting fatigue and power trade off would be affected by both the weights used in the optimization function, and loading component chosen.  

\begin{figure}
    \centering
    \includegraphics[width=0.8\textwidth]{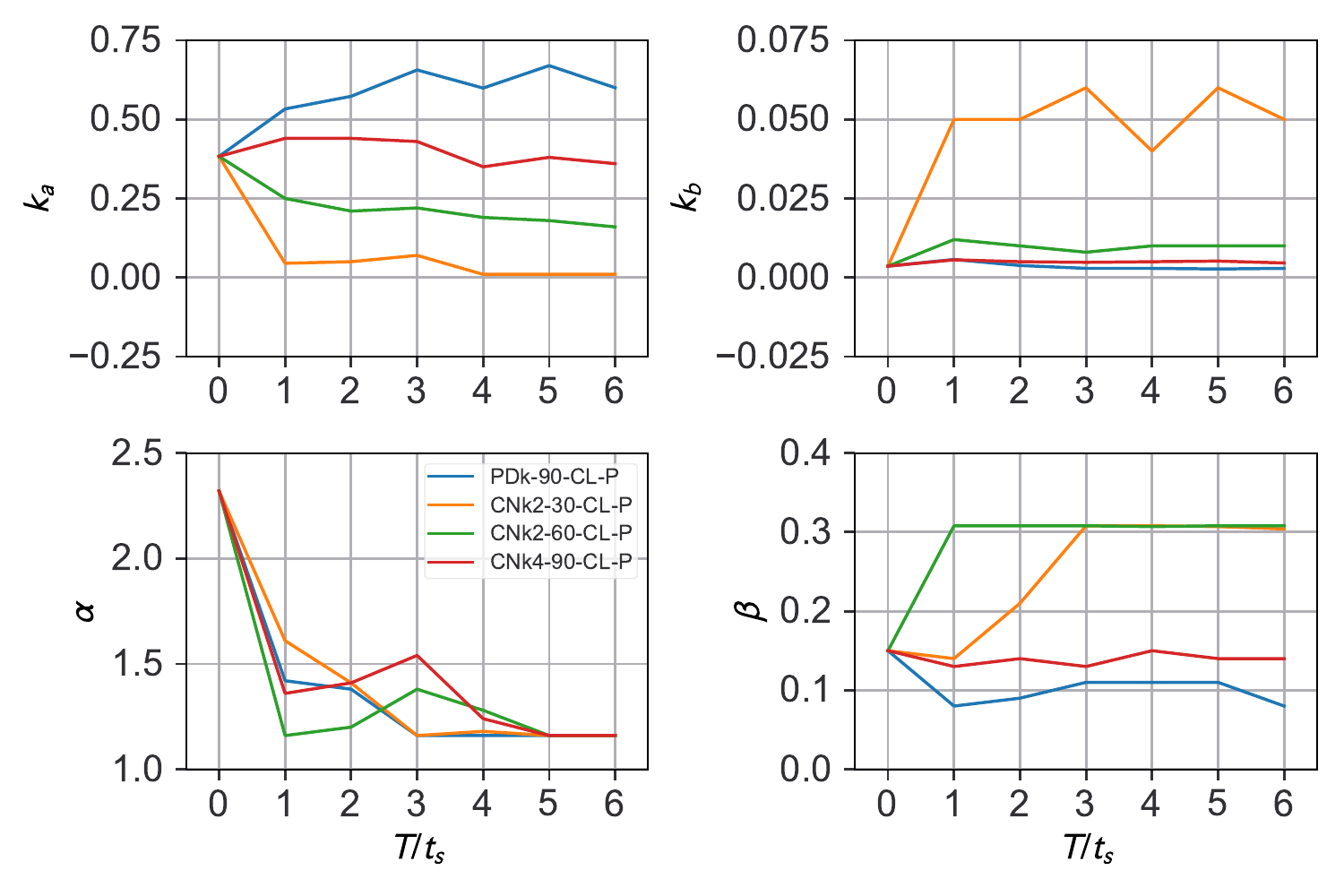}
    \caption[Evolution of the four wake model parameters as a result of online calibration.]{Evolution of the four wake model parameters as a result of online calibration. Results are presented for the closed-loop power optimization cases.}
    \label{fig:cali_time_series}
\end{figure}

To determine the effect of model calibration, the evolution of the four wake model parameters is shown in Figure~\ref{fig:cali_time_series}. Different combinations of the wake model parameters are obtained for each of the closed-loop power optimization case, further highlighting the sensitivity of the wake model to operating conditions of the farm. Similar to the yaw angles, the wake model parameters also appear to achieve statistically stable values. Longer simulations would be beneficial to confirm this, however the simulation time was restricted due to the high computational costs.

\section{Quasi-static control operational use case}

In the previous section, the benefits of the quasi-static closed-loop controller were shown through improvement in overall wind farm power production. The main effect driving the gains was the calibration of the wake model parameters, which resulted in better prediction and optimization of the farm performance for different inflow conditions. However, the operating conditions for all the case studied so far were statistically stable in terms of wind speed, wind direction and turbulence intensity. While the inflow conditions for the CNBL simulations may change with time, the current simulation length of \SI{3600}{\second} is not sufficient to cause significant changes in the inflow conditions. \cite{howland_cnbl} Hence, dynamically changing operational conditions are required to fully showcase the benefit of the developed closed-loop framework when compared to open-loop control. 

\begin{figure}
    \centering
    \includegraphics[width=0.5\textwidth]{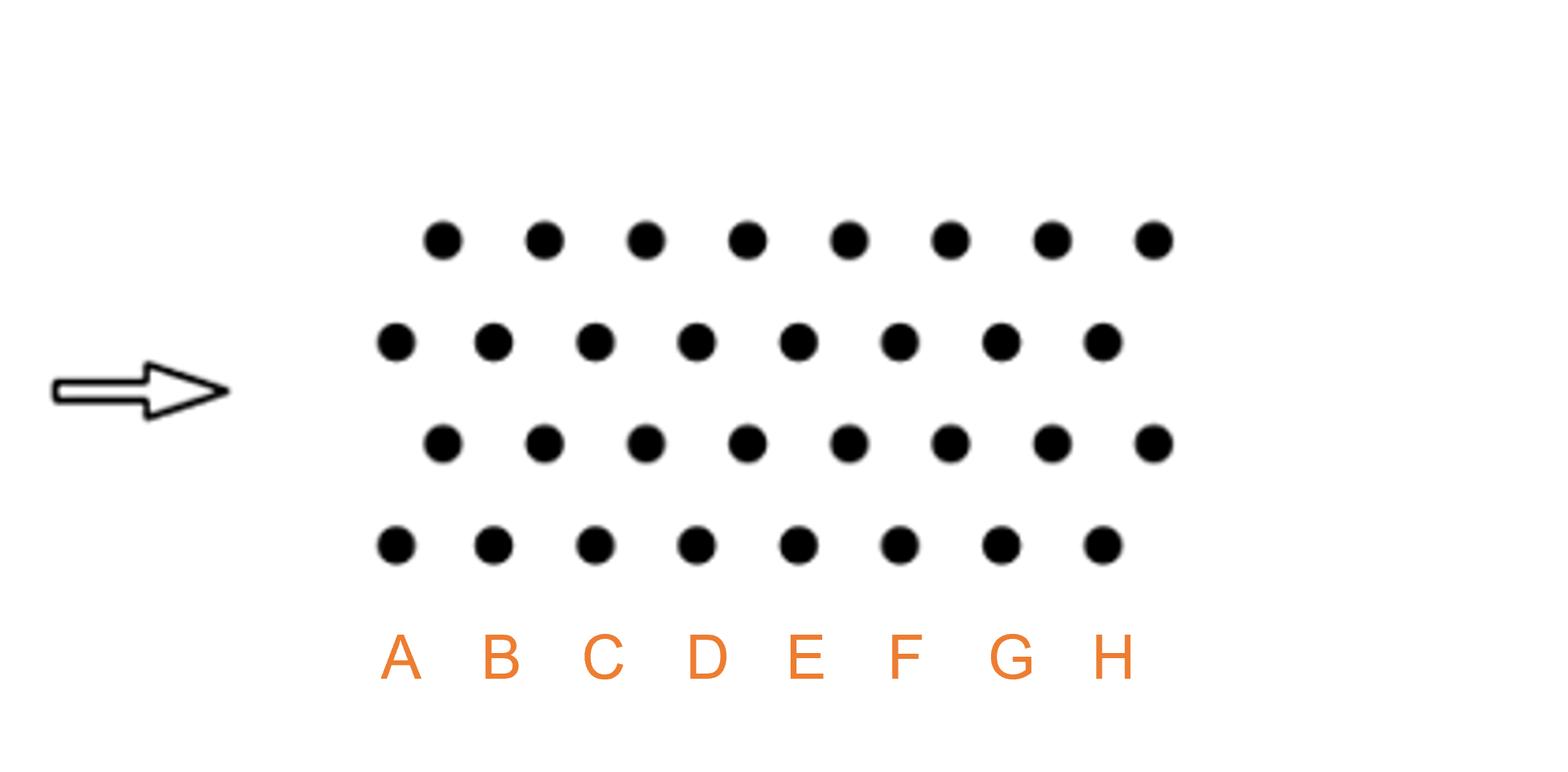}
    \caption[Fully aligned configuration of the TC RWP.]{Fully aligned configuration of the TC RWP. Four turbines are grouped per row, which are numbered A-H.}
    \label{fig:tc_rwp_rows}
\end{figure}

While previously studies have focused on developing closed-loop controllers for varying inflow conditions,\cite{Doekemeijer2020} the impact of operational scenarios such as turbine failure or downtime due to maintenance on coordinated wind farm flow control remains mostly unexplored. To this end, the quasi-static closed-loop control framework is used to study the impact of turbine shutdown on farm performance. A fully aligned configuration of the TC RWP, shown in Figure \ref{fig:tc_rwp_rows}, is used and six different scenarios defined, with three successive staggered rows of the farm being turned off. The benefit of the closed-loop control is quantified by comparing wind farm output against the same cases employing open-loop control. As open-loop control does not have feedback, it does not react to the changing wind turbine operation states, leading to improved performance gains by the closed-loop controller which continuously adapts the turbine yaw set points. Using the PDk inflow and focusing on both the power maximization and load mitigation objectives, the studied cases are listed in Table \ref{tab:opr_cases}.

\begin{table}
    \caption{Specifications of the operational power optimization case study. OL referes to open-loop control and CL to closed-loop control.}
    \centering
    \begin{tabular}{lcc}
    \hline
    Case & Row turned off & Control logic\\
    \hline
    \textit{PDk-90-OL-B}&B&OL P\\
    \textit{PDk-90-CL-B}&B&CL P\\
    \textit{PDk-90-CL-L-B}&B&CL P+L\\
    \textit{PDk-90-OL-C}&C&OL P\\
    \textit{PDk-90-CL-C}&C&OL P\\
    \textit{PDk-90-CL-L-C}&C&CL P+L\\
    \textit{PDk-90-OL-D}&D&OL P\\
    \textit{PDk-90-CL-D}&D&CL P\\
    \textit{PDk-90-CL-L-D}&D&CL P+L\\
    \hline
         &  \\
         & 
    \end{tabular}
    \label{tab:opr_cases}
\end{table}

A slight modification is necessary in the closed-loop framework to determine which turbines have been turned off, and hence should be included in the information loop shown in Figure \ref{fig:closed_loop_formulation} for the estimation, calibration or optimization steps. This is achieved by introducing a Boolean operator $B$ per turbine in equation \ref{eq:pgain} as follows

\begin{equation}
    P_{\textit{GAIN}} = \sum_{k=1}^{N_t} B_k \frac{C_P (\gamma_k) S_{k}^3 (\mathbf{\gamma})}{C_P (0) S_{k}^3 (0)}
\end{equation}

The Boolean operator takes on the values 0 or 1 for a turbine k based on whether the turbine had non-zero data points greater than a certain threshold in the last sampling period. In this study, we define a turbine to be off if the data was missing for more than 10\% of the last period, but different threshold values could be chosen based on the preference of the farm operator. The same Boolean operator can be used for DEL gains as well.

\begin{figure}
    \centering
    \includegraphics[width=\textwidth]{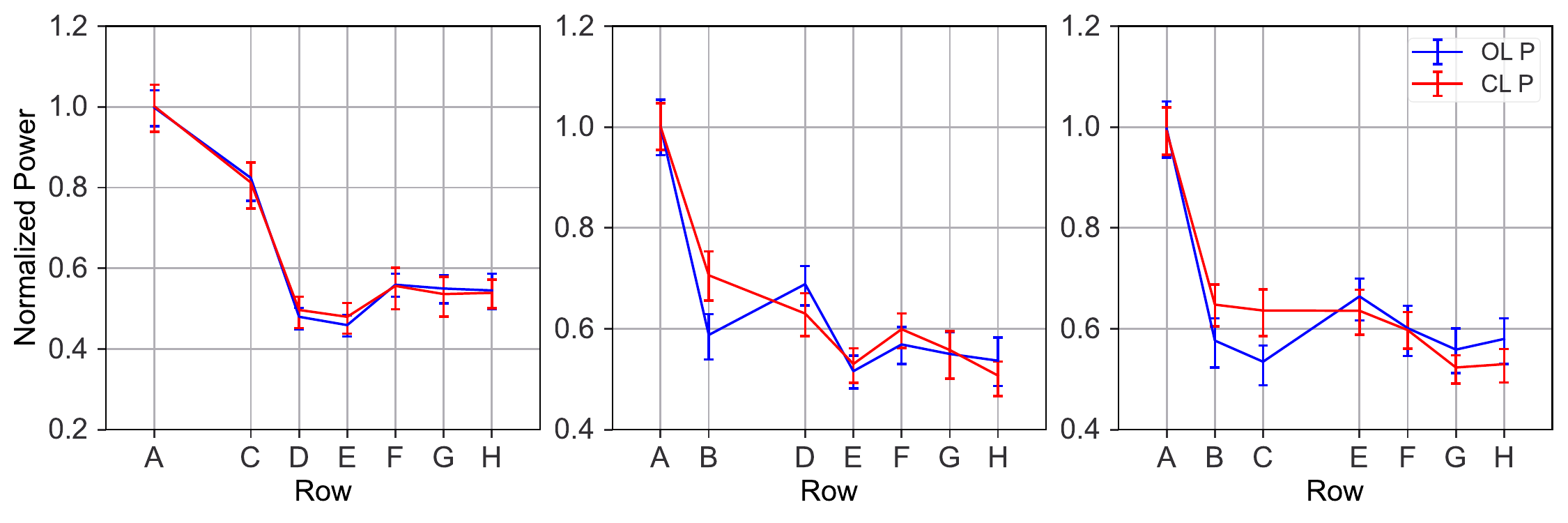}
    \caption[Effect of turning a turbine row off on row averaged power.]{Effect of turning a turbine row off on row averaged power. Comparison is made between open-loop and quasi static closed-loop control for power optimization.}
    \label{fig:opr_ind_power}
\end{figure}

\begin{figure}
    \centering
    \includegraphics[width=0.7\textwidth]{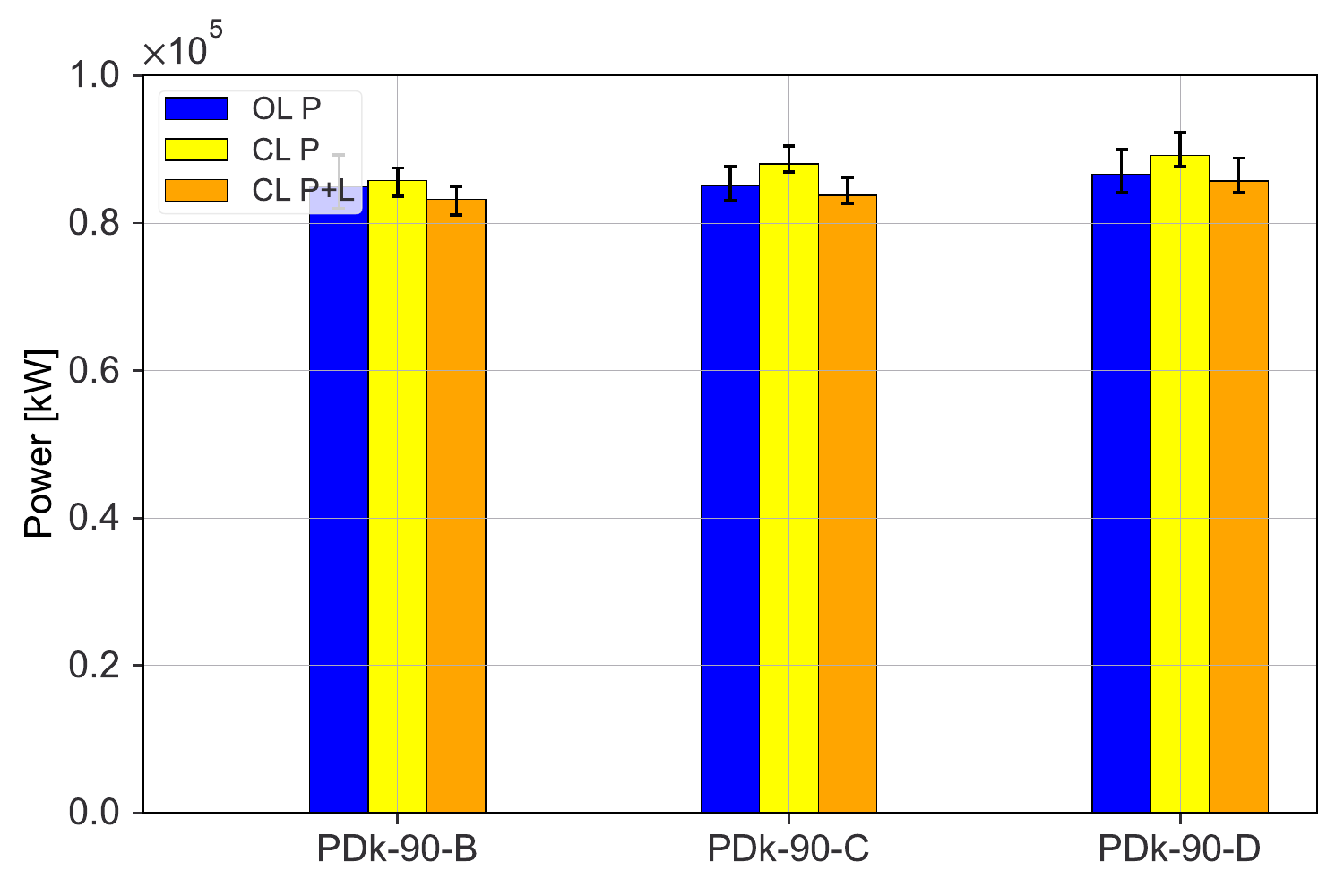}
    \caption[Effect of turning a turbine row off on total farm power.]{Effect of turning a turbine row off on total farm power. Comparison is made between open-loop and quasi static closed-loop control. Significant increase in power gains are obtained only for the case C and D, determined using a Welch's t-test.}
    \label{fig:opr_tot_power}
\end{figure}

The row averaged power for the operational cases is shown in Figure~\ref{fig:opr_ind_power} and the effect of the individual turbine gains on farm power can be seen in Figure~\ref{fig:opr_tot_power}, where we observe an increase in total farm power production, with significant gains visible for some of the turbine rows. However when looking at the overall wind farm power gains, the gains do not appear to be significant due to the overlapping error bars obtained via bootstrapping. To quantify the significance, a Welch's t-test is performed on the total wind farm power prediction between the open-loop and closed-loop gains. Amongst the three case pairs, the closed-loop mean power gains were significantly greater ($p<0.05$) than the open-loop gains for the cases PDk-90-C and PDk-90-D, while the PDk-90-B case did not have a statistically significant deference between the open-loop and closed-loop scenario. This indicates that the closed-loop control benefits are only evident when deeper turbine rows within the wind farm are turned off. Inclusions of loads in the optimization function lead to similar results as the previous sections, in which the closed-loop control power gains are similar to the open-loop gains.

\begin{figure}
    \centering
    \includegraphics[width=\textwidth]{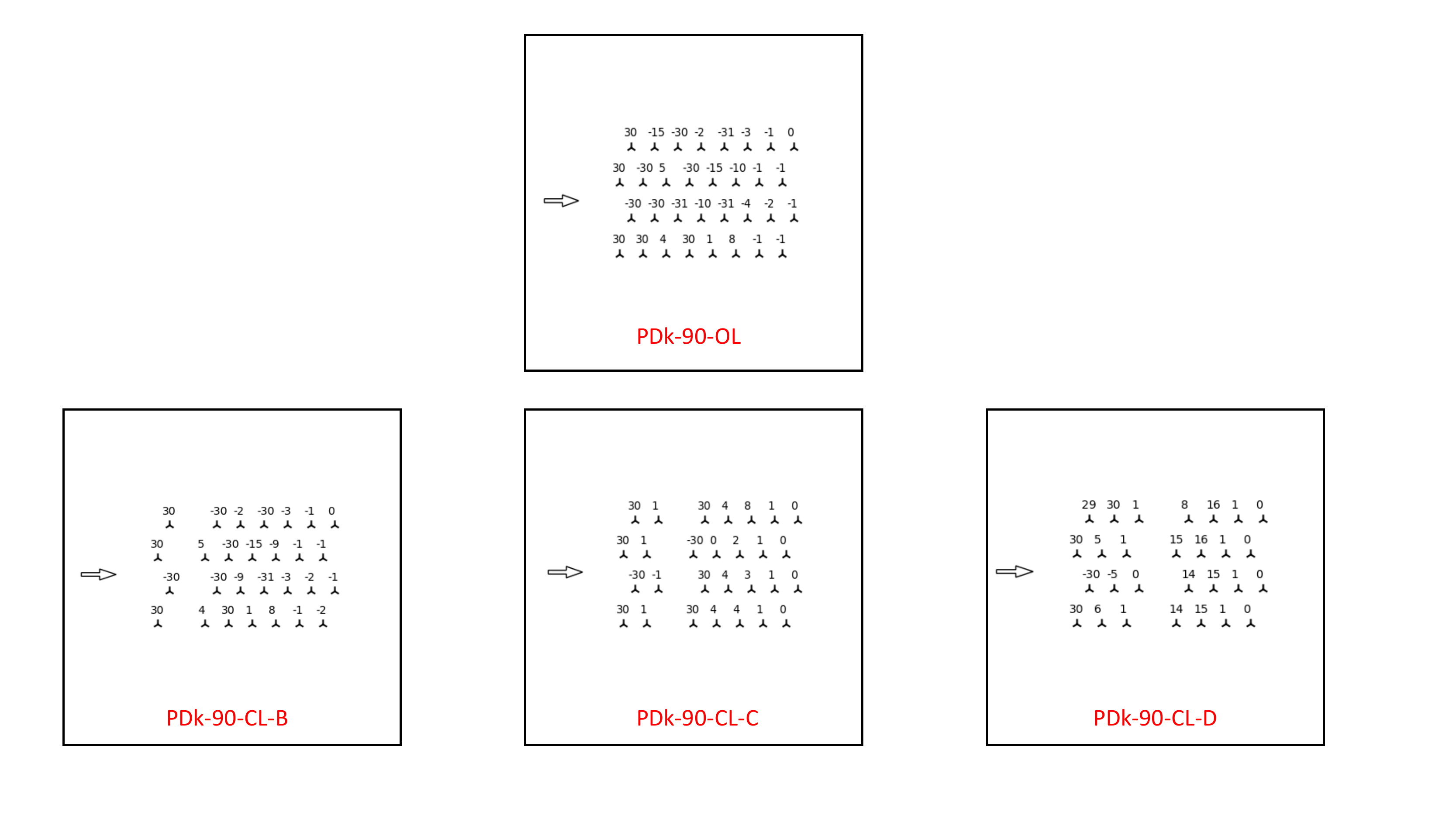}
    \caption[Effect of turning a turbine row off on optimal yaw set points.]{Effect of turning a turbine row off on optimal yaw set points. Comparison is made between open-loop and quasi static closed-loop control for power maximization.}
    \label{fig:opr_setpoints}
\end{figure}

The resulting yaw set points at the end of simulations are shown in Figure~\ref{fig:opr_setpoints}. From the optimal yaw set points, it is interesting to see that for cases PDk-90-C and PDk-90-D that the optimal yaw angles for the turbine right in front of the gap in the farm tends to be negligible. This is not the case for PDk-90-B, where the first turbine in each column is yawed by the maximum allowed angle, similar to the open-loop case. This can be attributed to higher wind speeds available to the front most turbines, which are subjected to free stream conditions. On the other hand for cases C and D, the turbines are operating in a reduced inflow condition and the power gain obtained by downstream turbines from wake steering away does not compensate for the loss of power.  In addition to the yaw angles, the wake recovery parameter for the CL cases also change, resulting in values which are similar to the ones obtained in Figure~\ref{fig:cali_time_series}.

\section{Conclusions, Discussion and Future work}

In this work, a combined power and fatigue optimization framework is developed using a quasi-static closed-loop controller through coordinated wake steering. A recursive wake model is used to estimate the wind farm power output, while a fatigue load LUT is used to estimate the increase in turbine fatigue due to wake steering. The load LUT was developed offline for the DTU \SI{10}{\mega\watt} reference turbine, and validated against reference LES simulations with and without wake steering. For purely power maximization, the developed closed-loop controller outperformed open-loop results obtained when tested in an aeroelastic LES based wind farm emulator. The gains were more pronounced for wind farm layouts with deep arrays, highlighting the benefit of online model calibration in wind direction scenarios leading to aligned turbines. Inclusion of fatigue in the optimization cost function through the developed load LUT lead to a reduction in both power produced and turbine blade root fatigue, but the power gains were still comparable to or greater than open-loop control. The balance of power and fatigue improvement however is dependent on the chosen weights in the optimization cost function. 

A case study is also performed which showcase the usefulness of the developed controller and load LUT in real world scenarios. The quasi-static controller is shown to react to wind turbines turning off in the field, which could be due to maintenance or failure. Contradictory to an open-loop controller which does not react to such changes in operation, the closed-loop controller lead to changes in coordinated wake steering set points across the farm, leading to improved power generation. 

Based on the observations, several avenues for future work can be identified. The frequency of the controller update in the current quasi-static framework was limited by the length of the simulation time due to computational resources. Future work could focus on longer simulation runs and identifying the ideal controller update time based on farm orientation and atmospheric conditions. A number of improvements in the wake model could also be made to improve its accuracy. Currently, impacts of atmospheric effects such as veer, shear, stability and Coriolis effects on turbine wake and recovery were not included in the wake model when determining the optimal set points. Furthermore, the model also lacked the influence of the counter rotating vortexes behind yawed turbines, which can result in apparent yawing of downstream turbines through the phenomenon known as secondary steering.~\cite{King2020} Inclusion of the aforementioned effects in the wake model used for control could lead to differences in the obtained optimal set points, and also reduce the prediction errors and calibration effort. Additional estimators for ambient conditions such as wind direction and turbulence intensity could also be included in the framework to address the estimation and observability challenges associated with closed-loop control. The wake model could also be made dynamic, similar to the FLORIDyn framework,~\cite{floridyn} to be able to react to fast changing scenarios in wind farm control and account for the advection of control effects throughout the farm at faster time scales than the ones considered in this work through the quasi-static assumption. To determine the optimal weights for the combined power and fatigue optimization, the financial impact of increased or reduced fatigue could also be assessed in future work, so that it can be balanced against changing day ahead electricity prices.

\section*{Acknowledgements} The authors have received funding from the European Unions Horizon 2020 programme (TotalControl, grant no. 727680 ). The computational resources and services used in this work were provided by the VSC (Flemish Supercomputer Center), funded by the Research Foundation Flanders (FWO) and the Flemish Government department EWI.

\subsection*{Author contributions}
IS and JM jointly defined the scope of the study. IS developed the closed-loop control methodology, implemented necessary algorithms and performed all the LES simulations. CE developed the fatigue load-look up table and performed the loads verification against the LES simulations of IS. The manuscript was written by IS, JM, and CE.

\subsection*{Conflict of interest}
The authors declare no potential conflict of interests.

\appendix
\section{Range of remaining input parameters }{\label{app1}}

The following sections illustrate the influence of the different input parameters of the load LUT on the resulting fatigue loads.

\subsection{Pitch angle offset}
\begin{figure}
	\centering
	\includegraphics[width=0.4\textwidth]{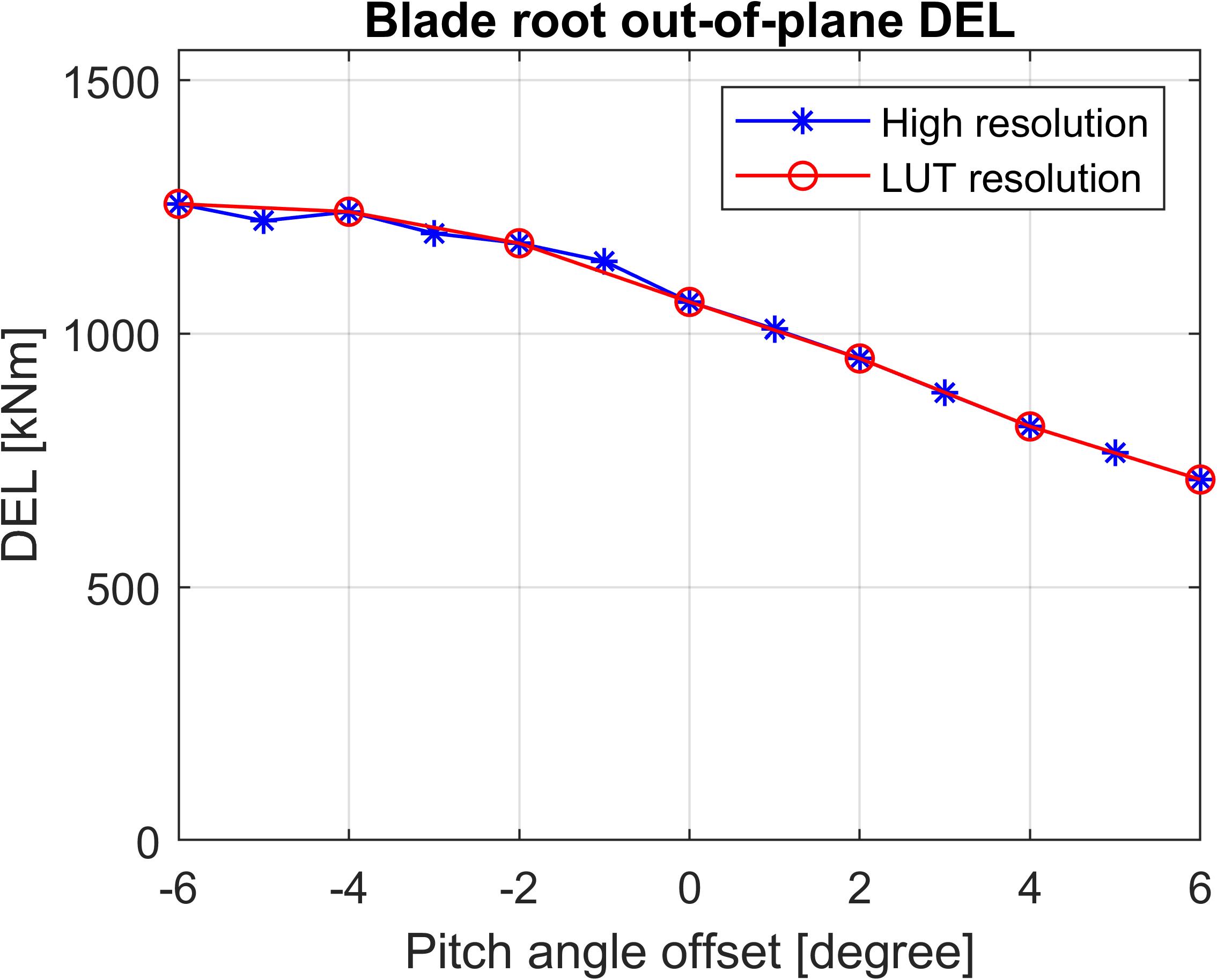}
	\caption{Blade root out-of-plane damage equivalent load as a function of blade pitch ($\beta$). The free-stream wind velocity ($U_\infty$) is set equal to 8 \SI{}{ms^{-1}} and the wake deficit depth $W_d$ is zero.}
	\label{fig:RootMOoP_DEL_po}
\end{figure}

Although pitch angle offset optimization in the context of induction control is not a part of this work, the input parameter is added to the load LUT to allow for combiend wake steering and induction control in future work. Extreme values of -6 and 6 degrees are chosen, so that both over-induction and under-induction can be applied. Figure \ref{fig:RootMOoP_DEL_po} shows a more or less linear pattern of the blade root out-of-plane DEL.

\subsection{Wake deficit depth}
The effect of wakes impinging on the controlled turbine a wind farm can be approximated by assuming the inflow wake to have a Gaussian wake shape as described by Niayifar and Port{\'e}-Agel. \cite{Niayifar2016} The Gaussian wake shape can be defined in terms of three input parameters used in the load LUT: the wake deficit depth $W_d$, the wake width $\sigma_D$ and the wake center location $\delta$. 
\begin{equation}\label{eq:Gaussian_wake_profile}
	\langle U\rangle = U_{\infty} \left(1-W_d \exp\left(-\frac{2 \left(y-\delta\right)^2}{D^2 \sigma_D^2}\right) \exp\left(-\frac{2 \left(z-z_h\right)^2}{D^2 \sigma_D^2}\right)\right)
\end{equation}

In Equation \ref{eq:Gaussian_wake_profile} $U_{\infty}$ is the free-stream flow velocity, $D$ the diameter of the rotor, $y$ the coordinate in the horizontal direction perpendicular to the flow direction (equal to zero at the hub of the turbine), $z$ the coordinate in the vertical direction (equal to zero at ground level) and $z_h$ the hub height. The wake width $\sigma_D$ is equal to $2\sigma/D$ with $\sigma$ the spread of the wake. For simplicity a symmetrical wake is used in this work. To simulate the effect of incoming wakes on a turbine in OpenFast, the wake shapes are added to the turbulence input files.

The wake deficit depth is defined as the maximum velocity deficit in the wake divided by the free-stream flow velocity. For the wake deficit depth the first extreme is zero, which means no wake is present and also the parameters wake width and center location become irrelevant. The second extreme value is 0.55, which is a value higher than the maximum observed in the TCRWP. The DEL trend is quasi-linear with a bend at a wake depth of about 0.3. This can be observed for the blade root out-of-plane DEL in Figure \ref{fig:RootMOoP_DEL_wd}. 

\subsection{Wake width}
Based on simulations on the TCRWP, the extreme values for the wake width are chosen to be 0.65 and 1.73. As can be observed in Figure \ref{fig:RootMOoP_DEL_ww} the blade root out-of-plane DEL is not influenced substantially by a variation in wake width. The quasi-linear trend can be approximated by using one extra value, namely 1.2 for the wake width. In total this input parameter is represented by three values in the load LUT.

\subsection{Wake center location}
\begin{figure}
	\centering
	\subfloat{\includegraphics[width=0.32\textwidth]{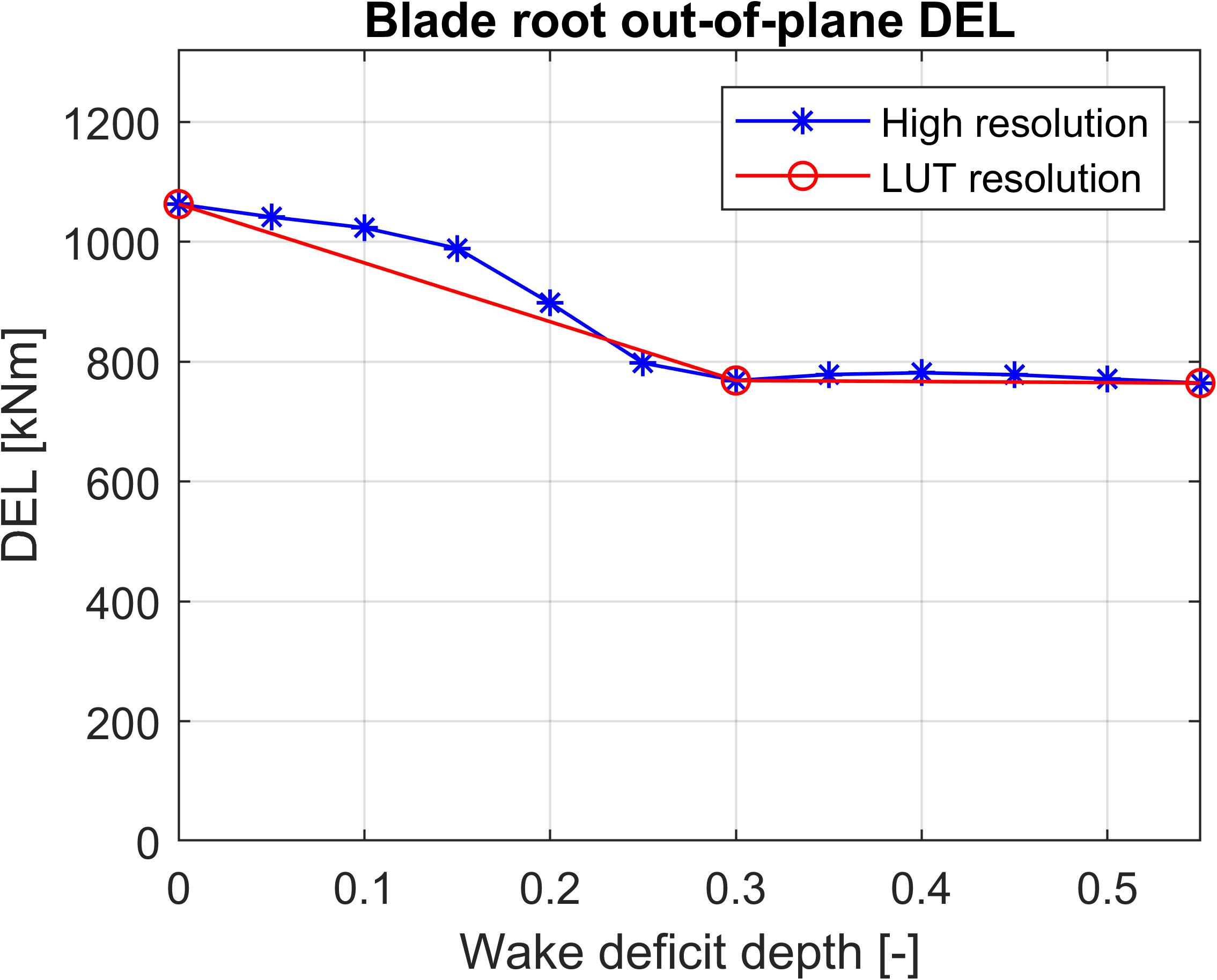}\label{fig:RootMOoP_DEL_wd}}
	\hspace{5pt}
	\subfloat{\includegraphics[width=0.32\textwidth]{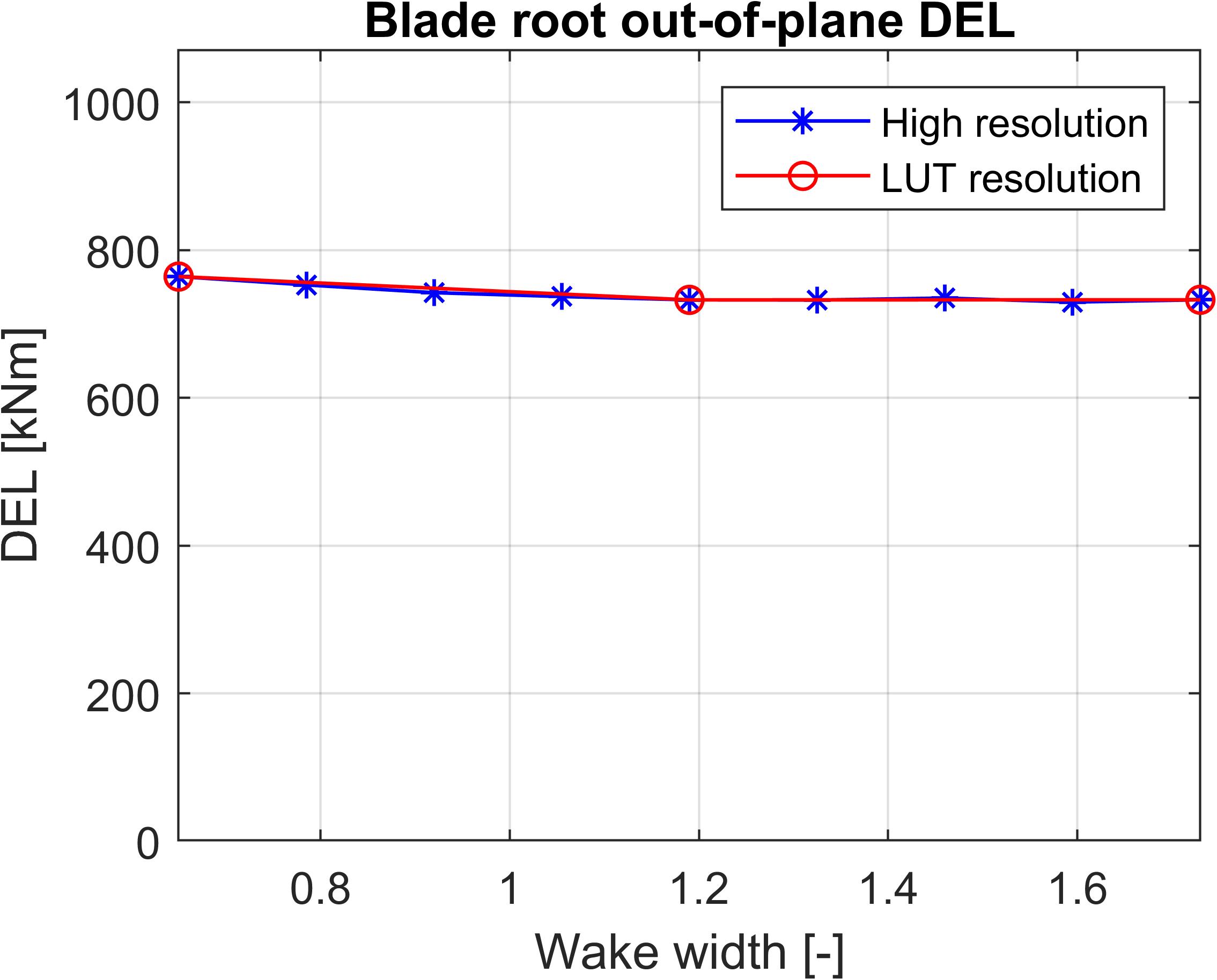}\label{fig:RootMOoP_DEL_ww}}
        \hspace{5pt}
	\subfloat{\includegraphics[width=0.32\textwidth]{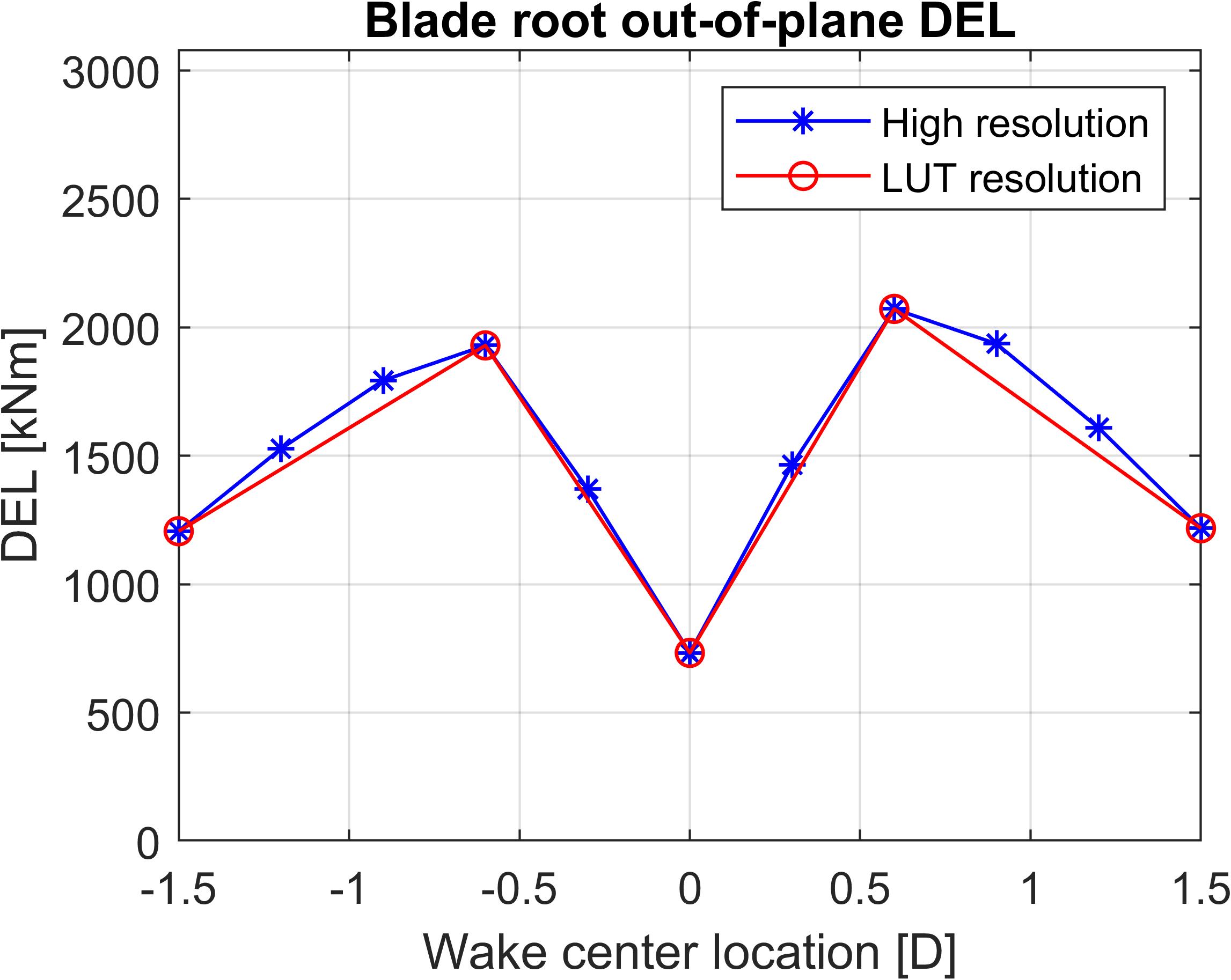}\label{fig:RootMOoP_DEL_wl}}
	\caption[Blade root out-of-plane damage equivalent load 1D analysis.]{Blade root out-of-plane damage equivalent load as a function of inflow wake parameters,(a) wake deficit depth ($W_d$), (b) wake width ($\sigma_D$) and (c) wake center location ($\delta$). The free-stream wind velocity ($U_\infty$) for all cases is set equal to 8 m/s.}
	\label{fig:DEL_RootMOoP_part3}
\end{figure}

This parameter is used to determine the center of the upstream wake with respect to the turbine for which the fatigue loads are being evaluated. By including this parameter, partial waked conditions can be represented in OpenFast. Representing the wake location in terms of rotor diameters with respect to the downstream turbines, the extreme values -1.5 D and 1.5 D are chosen for the wake center location, based on TCRWP simulations. Figure \ref{fig:RootMOoP_DEL_wl} shows that the blade root out-of-plane DEL has an M-shaped trend for the wake center location. The peaks of the M-shape at certain wake center locations are most likely caused by the uneven loading due to a highly uneven wind inflow velocity distribution during partial waked conditions. If the wake center location gets even further, the wake deficit depth is not seen anymore by the wind turbine and the DELs decrease again. At least five values for the wake center location are needed to form an M-shape. Therefore, the values $\left[-1.5, -0.6, 0, 0.6, 1.5\right]$ D are used to create the load LUT.

\section{Turbulence seed convergence study}{\label{app2}}

\begin{figure}
	\centering
	\includegraphics[width=0.7\textwidth]{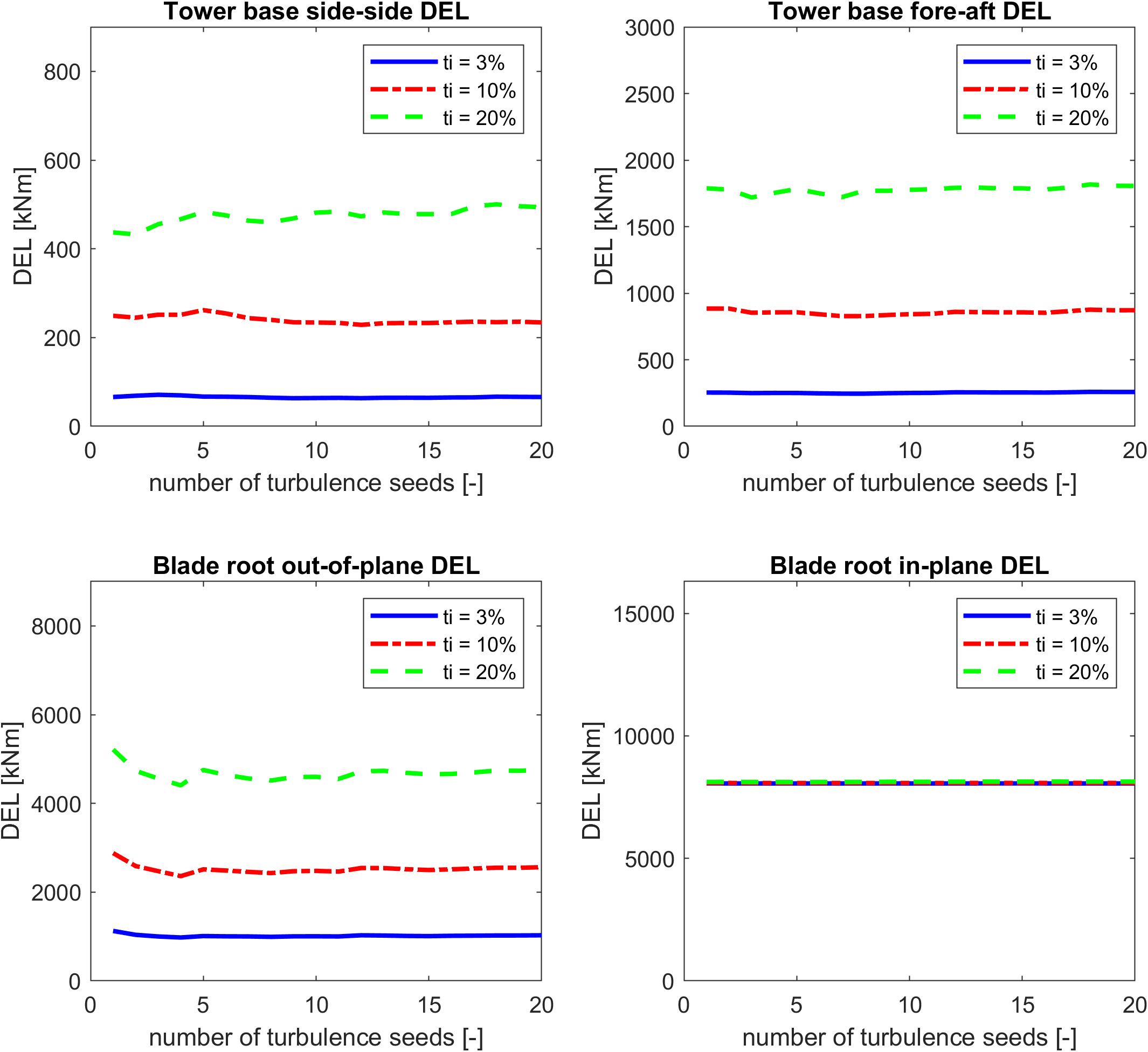}
	\caption[Damage equivalent load convergence against number of turbulence seeds.]{Damage equivalent load convergence against number of turbulence seeds.}
	\label{fig:TC}
\end{figure}

Figure \ref{fig:TC} shows the convergence of the different DELs for an increasing amount of turbulence seeds used. As expected from theory, the convergence is slower for higher turbulence intensities. Hence, higher $I$ values require more turbulence seeds. 

\end{document}